\documentclass[preprint,amsmath,amssymb]{revtex4-1}

\usepackage{graphicx}% Include figure files
\usepackage{bm}

\begin{document}

\title{A small-world of weak ties provides optimal global integration
  of self-similar modules in functional brain networks}

\author{Lazaros K. Gallos}
\affiliation{Levich Institute and Physics Department, City College of New York, New York, NY 10031, USA}
\author{Hern\'an A. Makse}
\affiliation{Levich Institute and Physics Department, City College of New York, New York, NY 10031, USA}
\author{Mariano Sigman}
\affiliation{Integrative Neuroscience Laboratory, Physics Department, FCEyN, Universidad de
Buenos Aires, Buenos Aires, Argentina}

\begin{abstract}

  The human brain is organized in functional modules. Such an
  organization presents a basic conundrum: modules ought to be
  sufficiently independent to guarantee functional specialization and
  sufficiently connected to bind multiple processors for efficient
  information transfer.  It is commonly accepted that small-world
  architecture of short lengths and large local clustering may solve
  this problem.  However, there is intrinsic tension between shortcuts
  generating small-worlds and the persistence of modularity; a global
  property unrelated to local clustering. Here, we present a possible
  solution to this puzzle. We first show that a modified percolation
  theory can define a set of hierarchically organized modules made of
  strong links in functional brain networks. These modules are
  ``large-world'' self-similar structures and, therefore, are far from
  being small-world. However, incorporating 
  weaker ties
  to the
  network converts it into a small-world preserving an underlying
  backbone of well-defined modules. Remarkably, weak ties are
  precisely organized as predicted by theory maximizing information
  transfer with minimal wiring cost. This trade-off architecture is
  reminiscent of the ``strength of weak ties'' crucial concept of
  social networks. Such a design suggests a natural solution to the
  paradox of efficient information flow in the highly modular
  structure of the brain.
\end{abstract}

\maketitle

\section{Introduction}

One of the main findings in Neuroscience is the modular organization
of the brain which in turn implies the parallel nature of brain
computations \cite{dehaene2001towards,meunier,bassett}.  For example,
in the visual modality, more than thirty visual areas analyze
simultaneously distinct features of the visual scene: motion, color,
orientation, space, form, luminance and contrast among others
\cite{felleman1991distributed}. These features, as well as information
from different sensory modalities, have to be integrated, as one of
the main aspects of perception is its unitary nature
\cite{dehaene2001towards,treisman1996binding}.

This leads to a basic conundrum of brain networks: modular processors
have to be sufficiently isolated to achieve independent computations,
but also globally connected to be integrated in coherent functions
\cite{dehaene2001towards,meunier,tononi1994measure}.  A 
current view is that small-world networks provide a solution to this
puzzle
since they combine high local clustering and short path length
\cite{Sporns2004,watts98,sw}.  This view has been fueled by the
systematic finding of small-world topology in a wide range of human
brain networks derived from structural \cite{he2007small}, functional
\cite{eguiluz2005scale,achard2006resilient,achard2007efficiency}, and
diffusion tensor MRI \cite{hagmann2007mapping}.  Small-world topology
has also been identified at the cellular-network scale in functional
cortical neuronal circuits in mammals
\cite{song2005highly,yu2008small} and even in the nervous system of
the nematode {\it Caenorhabditis elegans} \cite{watts98}.
Moreover, small-world property seems to be relevant for brain function
since it is affected by disease \cite{stam2007small}, normal aging and
by pharmacological blockade of dopamine neurotransmission
\cite{achard2007efficiency}.

While brain networks show  small-world properties, 
several experimental studies have also shown that they
are hierarchical, fractal and highly modular
\cite{meunier,bassett,hierarchical,nested}.  As there is an intrinsic
tension between modular and small-world organization, the main aim of
this study is to reconcile these ubiquitous and seemingly
contradictory topological properties.
Indeed, traditional models of small-world networks cannot fully
capture the coexistence of highly modular structure with broad global
integration.
First, clustering is a purely local quantity which can be assessed
inspecting the immediate neighborhood of a node \cite{watts98}. On the
contrary, modularity is a global property of the network, determined
by the existence of strongly connected groups of nodes that are only
loosely connected to the rest of the network
\cite{newman1,fortunato,meunier,bassett}.
In fact, it is easy to construct modular and unclustered networks or,
reciprocally, clustered networks without modules.

Second, the short distances of a small-world may be incompatible with
strong modularity which typically presents the properties of a
``large-world''
\cite{song05,song06,goh06,radicchi08,rozenfeld,lazaros,galvao}
characterized by long distances which effectively suppress diffusion
and free flow in the system \cite{lazaros}.  While a clustered network
preserves its clustering coefficient when a small fraction of
shortcuts are added (converting it into a small-world) \cite{watts98},
the persistence of modules is not equally robust.
As we show below, shrinking the network diameter may quickly destroy
the modules.

Hence, the concept of small-world may not be
entirely sufficient to explain the modular and integration features of
functional brain networks on its own.  We propose that a solution to
modularity and broad integration can be achieved by a network
in which
strong links form
large-world fractal modules, in agreement with
\cite{meunier,bassett,hierarchical,nested}, which are short-cutted by
weaker links establishing a small-word network.  A modified
percolation theory \cite{bunde-havlin,bollobas} can identify a
  sequence of critical values of connectivity thresholds forming a
  hierarchy of modules which progressively merge together. This
proposal is inspired by a fundamental notion of sociology termed by
Granovetter as ``the strength of weak ties''
\cite{granovetter,kertesz}. According to this theory, strong ties
(close friends) clump together forming modules. An acquaintance (weak
tie) becomes a crucial bridge (a shortcut) between the two densely
knit clumps (modules) of close friends \cite{granovetter}.

Interestingly, 
this theme also emerges in theoretical models of large-scale cognitive
architecture.
Several theories suggest integration mechanisms based on dynamic
binding \cite{tononi1994measure,tononi1998consciousness} or on a
workspace system \cite{dehaene2001towards,baars1997theater}.  For
instance, the workspace model
\cite{dehaene2001towards,baars1997theater} proposes that a flexible
routing system with dynamic and comparably weaker connections
transiently connects modules with very strong connections carved by
long-term learning mechanisms.

\section{Results}

\subsection{Experimental design and network construction}

We capitalize on a well-known dual-task paradigm, the psychological
refractory period \cite{Pashler}.
A total of 16 subjects 
responded with the right hand to a visual stimulus and with the left
hand to an auditory stimulus (see SI Appendix).  The temporal gap
between the auditory and visual stimuli varied in four stimulus onset
asynchrony, SOA= 0, 300, 900 and 1200 ms.  The sequence of activated
regions which unfolds during the execution of the task has been
reported in a previous study \cite{sigman2008brain}.

The network analysis relies on the time-resolved BOLD-fMRI response based
on the phase signal obtained for each voxel of data
\cite{sigman2007parsing}. 
We first compute the phase of the BOLD signal for each voxel with
methods developed previously \cite{sigman2007parsing}.  For each
subject and each SOA task, we obtain the phase signal of the $i$-th
voxel of activity, $\phi_i(t)_{\{t=1, .. , T\}}$, over $T=40$ trials
performed for a particular SOA value and subject.  We use these
signals to construct the network topology of brain voxels based on the
equal-time cross-correlation matrix, $C_{ij}$, where a network link
indicates a high cross-correlation in the phase activity of the two
voxels (see SI Appendix). The accuracy of the calculated $C_{ij}$
  values was estimated through a boot strapping analysis. The 95\%
  confidence interval becomes more narrow for higher $C_{ij}$ values,
  e.g., for $C_{ij}=0.975$ it is (0.9744, 0.9760). The corresponding
  standard deviation is of the order of 0.003. Thus, we typically
  distinguish between values that differ by 0.005 (see SI Appendix and
  Fig.~S1).

To construct the network, we link two voxels if their
cross-correlation $C_{ij}$ is larger than a predefined threshold value
$p$ \cite{eguiluz2005scale,achard2006resilient,salvador1}.  The
resulting network for a given $p$ is a representation of functional
relations among voxels for a specific subject and SOA.  We obtain 64
cross-correlation networks resulting from the four SOA values
presented to the 16 subjects.

\subsection{Percolation analysis}

Graph analyses of brain correlations
relies on a threshold 
\cite{eguiluz2005scale}
which is problematic since small-world like properties are sensitive
to even a small proportion of variation in the connections. The
present analysis
may be seen as an attempt to solve this problem.

The thresholding procedure can be naturally mapped to a percolation
process (defined in the $N\times N$ space of interactions $C_{ij}$);
a model to describe geometrical phase transitions of connected
clusters in random graphs, see Chapters 2 and 3 in \cite{bunde-havlin}
and \cite{bollobas,stanley}.

In general, the size of the largest component of connected nodes in a
percolation process remains very small for large $p$. The crucial
concept is that the largest connected component increases rapidly
through a critical phase transition at $p_c$, in which a single
incipient cluster dominates and spans the system
\cite{bunde-havlin,bollobas,stanley}.  A unique connected component is
expected to appear if the links in the network are occupied at random
without correlations.
However, when we apply the percolation analysis to the functional
brain network, a more complex picture emerges revealing a hierarchy of
clusters arising from the non-trivial correlations in brain activity.

\begin{figure}
\centerline{\resizebox{16.0cm}{!}  { \includegraphics{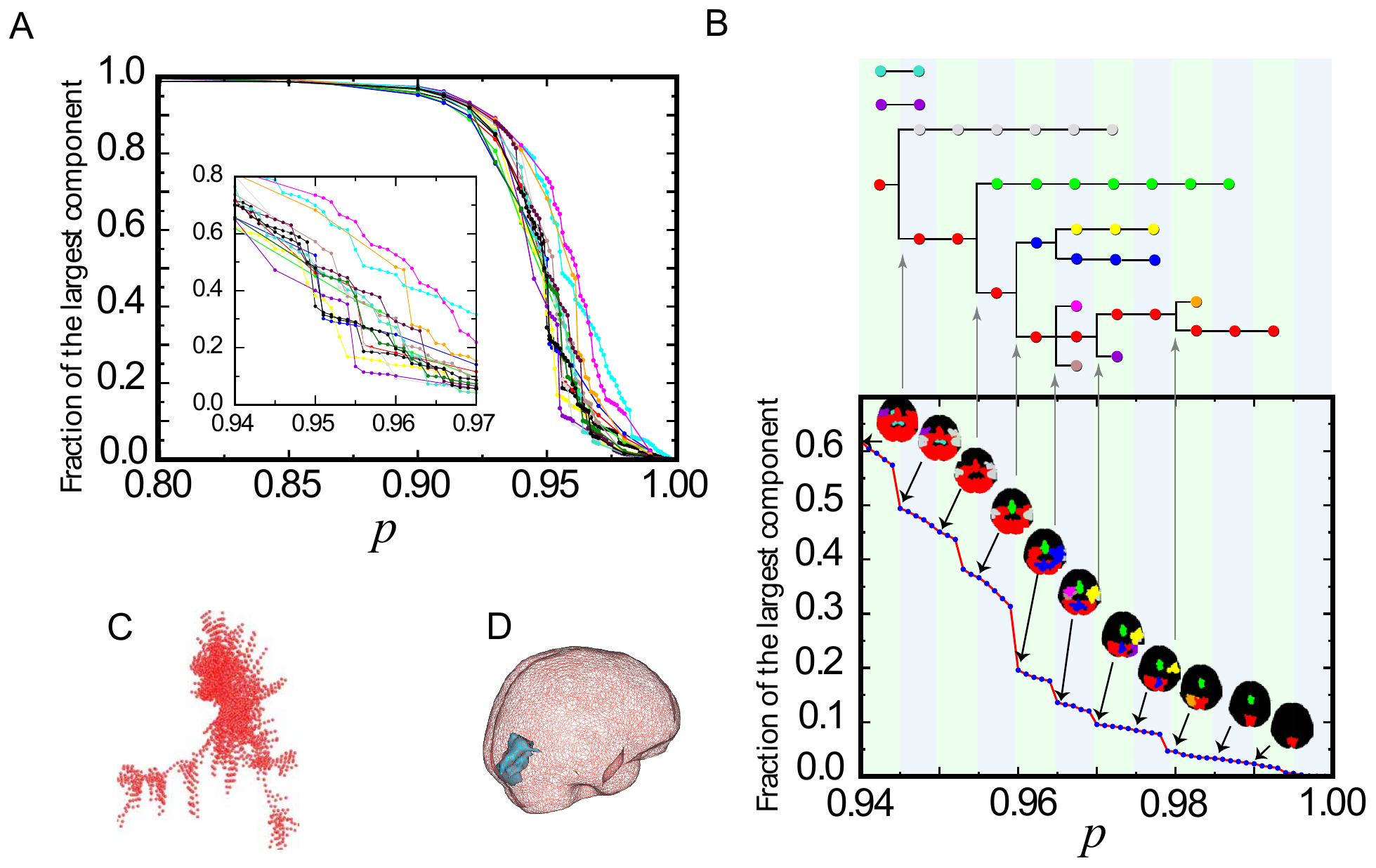}} }
\caption{  \label{perco}
{\bf Percolation analysis.}
{\bf (A)} Size of the
largest connected component of voxels (as measured by the fraction to
the total system size) versus $p$ for the 16 subjects 
(SOA=900 ms). The other three SOA values give similar results.
The inset presents a detail around $p\approx 0.95$.  {\bf (B)} Detail
for a representative  individual.  As we lower $p$ the size
of the largest component increases in jumps when new modules emerge,
grow, and finally get absorbed by the largest component.  We show the
evolution of the modules by plotting connected components with more
than 1,000 voxels. The hierarchical tree at the top of the plot
shows how clusters evolve by merging with each other. {\bf (C)} A typical module in network
representation.  {\bf (D) } The same module as in {\bf (C)} embedded
in real space - this specific module projects to the medial occipital
cortex, see SI Appendix 
for the spatial
projection of all modules.}
\end{figure}

For each participant, we calculate the size of the largest connected
component
as a function of $p$.  We find that the largest cluster size increases
progressively with a series of sharp jumps (Fig.~\ref{perco}A, SOA=900
ms, all participants, other stimuli are similar).  This suggests a
multiplicity of percolation transitions where percolating modules
subsequently merge as $p$ decreases rather than following the typical
uncorrelated percolation process with a single spanning cluster. Each
of these jumps defines a percolation transition focused on groups of
nodes which are highly correlated, constituting well-defined modules.

Figure \ref{perco}B shows the detailed behavior of the jumps in a
typical individual (subject labeled \#1 in our dataset \cite{dataset},
SOA=900 ms).
  At high values of $p$, three large clusters are formed localized
  to the medial occipital cortex (red), the lateral occipital cortex
  (orange) and the anterior cingulate (green).  At a lower $p=0.979$,
  the orange and red clusters merge as revealed by the first jump in
  the percolation picture.  As $p$ continues to decrease this
  mechanism of cluster formation and absorption repeats, defining a
  hierarchical process as depicted in the top panel of
  Fig. \ref{perco}B.  This analysis further reveals the existence of
  ``stubborn'' clusters.  For instance, the anterior cingulate cluster
  (green), known to be involved in cognitive control
  \cite{zylberberg2011human,duncan2010multiple} and which hence cannot
  commit to a specific functional module,
  remains detached from the other clusters down to low $p$ values.
  Even at the lower values of $p$, when a massive region of the cortex
  including motor, visual and auditory regions has formed a single
  incipient cluster (red, $p\approx 0.94$), two new clusters emerge;
  one involving subcortical structures including the thalamus and
  striatum (cyan) and the other involving the left frontal cortex
  (purple).  This mechanism reveals the iteration of a process by
  which modules form at a given $p$ value and merged by comparably
  weaker links.  This process is recursive. The weak links of a given
  transition become the strong links of the next transition, in a
  hierarchical fashion.

Here, we focus our analysis on the first jump in the size of the
largest connected component, for instance, $p_c=0.979$ in
Fig. \ref{perco}B.
We consider the three largest modules at $p_c$ with at least 1,000
voxels each. This analysis results in a total of 192 modules among all
participants and stimuli which are pooled together for the next
study. An example of an identified module in the medial occipital
cortex of subject \#1 and SOA=900 ms is shown in Fig. \ref{perco}C in
the network representation and in Fig. \ref{perco}D in real space.
The topography of the modules reflects coherent patterns across the
subjects and stimuli as analyzed in SI Appendix (see
Fig.~S2).

\subsection{ Scaling analysis and Renormalization Group}

To determine the structure of the modules we investigate the scaling
of the ``mass'' of each module (the total number of voxels in the
module, $N_c$) as a function of three length-scales defined for each
module: {\it (i)} the maximum path length, $\ell_{\rm max}$, {\it
  (ii)} the average path length between two nodes, $\langle \ell
\rangle$, and {\it (iii)} the maximum Euclidean distance among any two
nodes in the cluster, $r_{\rm max}$. The path length, $\ell$, is the
distance in network space defined as the number of links along the
shortest path between two nodes. The maximum $\ell_{\rm max}$ is the
largest shortest path in the network.

\begin{figure}
\centerline{\resizebox{16.0cm}{!} { \includegraphics{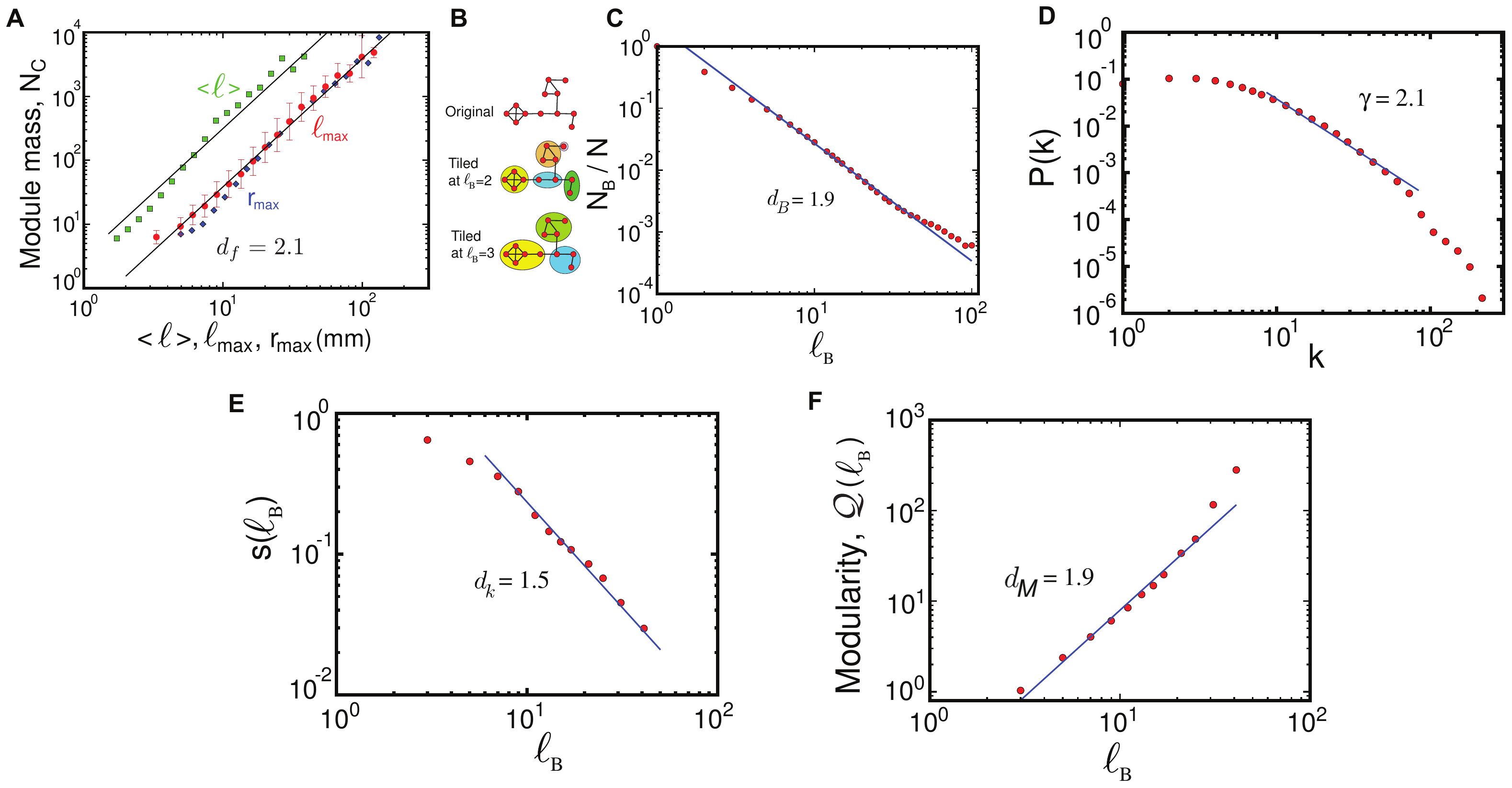}}}
\caption{\label{fractal} 
{\bf Strong ties define fractal modules}.
{\bf (A)}
Number of voxels or mass of each module, $N_c$, versus
$\ell_{\rm max}$ (red $\bullet$), $\langle \ell \rangle$ (green
$\blacksquare$), and $r_{\rm max}$ (blue $\blacklozenge$).  Each point
represents a bin average over the modules for all subjects and
stimuli. We use all the modules appearing at $p_c$.  The straight
lines are fittings according to Eq. (\ref{df}). The variance is the
statistical error over the different modules.  The variance is similar
in the other data.
{\bf (B)}
Detection of submodules and fractal dimension inside the percolation
modules.  We demonstrate the box-covering algorithm for a schematic
network, following the Maximum Excluded Mass Burning algorithm in
\cite{song05,jstat} (SI Appendix). 
We cover a network with boxes of size $\ell_B$ which are identified as
sub-modules.  We define $\ell_B$ as the shortest path plus one.
{\bf (C)}
Scaling of the number of boxes $N_B$ needed to cover the
network of a module versus $\ell_B$ yielding $d_B$.  We average over
all the identified modules for all subjects.
{\bf (D)}
Degree distribution averaged over all the brain modules.
The individual degree distributions for each module (Fig.~S4 and
SI Appendix) roughly follow a power law with an average
exponent $\gamma = 2.11\pm 0.04$.
{\bf (E)}
Dependence of the scaling factor $s(\ell_B)$, defined through 
$k'=s(\ell_B) k$ for the renormalized degree $k'$, on 
$\ell_B$. The exponent $d_k=1.5$  characterizes how the node degree
changes during the renormalization process.
{\bf (F)}
Quantification
of the modularity of the brain modules.  The identified percolation
modules at $p_c$ are composed of submodules with a high level of
modularity as can be seen by the scaling of ${\cal Q}(\ell_B)$ with
$\ell_B$ that yields a large modularity exponent $d_M=1.9\pm 0.1$.
Deviations from linear scaling are found at large $\ell_B$ due to
boundary effects since the network is reduced to just a few
submodules.
}
\end{figure}

Figure \ref{fractal}A indicates power-law scaling for
these quantities
\cite{song05,bunde-havlin}. For instance:
\begin{equation}
  N_c(r_{\rm max}) \sim (r_{\rm max})^{d_f},
\label{df}
\end{equation}
defines the Euclidean Hausdorff fractal dimension, $d_f = 2.1\pm0.1$.
The scaling with $\ell_{\rm max}$ and $\langle \ell \rangle$ is
consistent with Eq. (\ref{df}) as seen in Fig. \ref{fractal}A. The
exponent $d_f$ quantifies how densely the volume of the brain is
covered by a specific module.

Next, we investigate the network properties of each module,
applying Renormalization Group (RG) analysis for complex networks
\cite{song05,song06,goh06,radicchi08,rozenfeld}.  This technique
allows one to observe the network at different scales transforming it
into successively simpler copies of itself, which can be used to
detect characteristics which are difficult to identify at a specific
scale of observation.  We use this technique to characterize
sub-modular structure within each brain module \cite{meunier}.

We consider each module identified at $p_c$ separately. We then tile
it with the minimum number of boxes or sub-modules, $N_B$, of a given
box diameter $\ell_B$ \cite{song05}, i.e., every pair of nodes in a
box has shortest path length smaller than $\ell_B$. Notice that
  the calculations are performed in network space, where path lengths
  are defined across the network links without the need for an
  embedding dimension.

Covering the network with minimal $N_B$ sub-modules represents an
optimization problem which is solved using standard box-covering
algorithms, such as the Maximum Excluded Mass Burning algorithm,
  MEMB, which was introduced in \cite{song05,song06,jstat} to describe
  the self similarity of complex networks ranging from the WWW,
  biological and technical networks (see SI Appendix
  and Fig. \ref{fractal}B describing MEMB; the code can be downloaded
from \cite{dataset}).  The requirement to minimize the number of boxes
is important since the resulting boxes are characterized by the
proximity between all their nodes and minimization of the links
connecting the boxes \cite{lazaros}. Thus, the box-covering algorithm
detects boxes/submodules that also tend to maximize modularity.

The repetitive application of box-covering at different values of
$\ell_B$ is a RG transformation \cite{song05} that yields a different
partition of the brain modules in submodules of varying sizes
(Fig. \ref{fractal}B). Figure \ref{fractal}C shows the scaling of $N_B$ versus
$\ell_B$ averaged over all the modules for all individuals and
stimuli. This property is quantified in the power-law relation
\cite{song05}:

\begin{equation}
  N_B(\ell_B) \sim \ell_B^{-d_B}, 
\label{db}
\end{equation}
where $d_{\rm B}$ is the box fractal dimension
\cite{song05,song06,goh06,radicchi08,rozenfeld} which characterizes
the self-similarity between different scales of the module where
smaller-scale boxes behave in a similar way as the original network.
The resulting $d_B$ averaged over all the modules is $d_B=1.9 \pm 0.1$.

\subsection{Morphology of the brain modules}

  The RG analysis reveals that the module topology does not have
  many redundant links and it represents the quantitative statement
  that the brain modules are ``large-worlds''.  However this analysis
  is not sufficient to precisely characterize the topology of the
  modules.  For example, a two-dimensional complex network
  architecture and a
  simple two-dimensional lattice are compatible with the scaling
  analysis and the value of the exponents described in the previous
  section.

  To identify the network architecture of the modules
  we follow established analysis \cite{scalefree,hierarchical}
  based on the study of the degree distribution of the modules,
  $P(k)$, and the degree-degree correlation $P(k_1,k_2)$. The form of
  $P(k)$ distinguishes between a Euclidean lattice (delta function),
  an Erdos-Renyi network (Poisson) \cite{bollobas}, or a scale-free
  network (power-law) \cite{scalefree}.
%In Fig. \ref{fractal}D we find
%that the brain modules have a broad degree distribution
%\cite{scalefree,eguiluz2005scale} with an approximate power-law
%$P(k)\sim k^{-\gamma}$, where $\gamma = 2.2\pm 0.1$, for
%intermediate values of $k$, in agreement with results reported in
%the literature \cite{eguiluz2005scale}.
  We find that the brain modules have a broad degree distribution
  \cite{scalefree,eguiluz2005scale} with an approximate power-law
  $P(k)\sim k^{-\gamma}$. The statistical analysis provides strong
  evidence for a power law form and rules out exponential decay (see
  SI Appendix).
  In Fig.~S4 we present a number of $P(k)$ curves for different modules,
  along with their best fittings. In the SI Appendix we describe
  the calculation method that takes into account all the clusters
  and finally yields an average exponent $\gamma=2.11\pm 0.04$.
  An `average' curve for the distribution is shown in Fig.~\ref{fractal}D,
  where the exponent $\gamma$ is not a direct fit to this curve,
  but instead represents the result of the accurate calculation.
  This result indicates that
  the modules have a scale-free fractal structure, different from a simple
  two-dimensional lattice, where $P(k)$ should be a narrow function.

  The embedding of scale-free networks in a finite-dimension real
  space constitutes a problem which has attracted recent attention
  \cite{arcangelis,shlomo1,shlomo2}.  Scale-free networks may arise
  from a 2-dimensional lattice with added dense connectivity locally,
  where the weights and connectivity are inversely proportional to the
  Euclidean distance on the lattice.  To investigate this possibility
  we study the correlation function of the phases of the voxels as a
  function of Euclidean distance in real space: $ C(r) = \langle
  \cos(\phi_1-\phi_2) \rangle $ versus $r=|\vec{r}_1-\vec{r}_2|$.
  This function can be interpreted as the correlation between two
  spins with orientation determined by the phase $\phi_i$ of the voxel
  at location $\vec{r}_i$ (average is over all pairs at distance $r$).
  We find (see SI Appendix and Fig.~S3) that $C(r)$ decays
  algebraically with distance.  Thus, our results indicate that
  modules are scale-free networks which can be embedded in a lattice
  with an added long-range connectivity.

  How can fractal modularity emerge in light of the scale-free
  property, which is usually associated with small-worlds
  \cite{hierarchical}?  In a previous study \cite{song06}, we
  introduced a model to account for the simultaneous emergence of
  scale-free, fractality and modularity in real networks by a
  multiplicative process in the growth of the number of links, nodes
  and distances in the network. The dynamic follows the inverse of the
  RG transformation \cite{song06} where the hubs acquire new
  connections by linking preferentially with less connected nodes
  rather than other hubs.  This kind of ``repulsion between hubs''
  \cite{goh06} creates a disassortative structure--- with hubs
  spreading uniformly in the network and not crumpling in the core as
  in scale-free models \cite{scalefree}. Hubs are buried deep into the
  modules, while low degree nodes are the intermodule connectors
  \cite{goh06}.

  A signature of such mechanism can be found by following hubs' degree
  during the renormalization procedure.
  At scale $\ell_B$, the degree of a hub $k$ changes to the degree of
  its box $k^\prime$, through the relation $k^\prime=s(\ell_B) k$. The
  dependence of the scaling factor $s(\ell_B)$ on $\ell_B$ defines the
  renormalized degree exponent $d_k$ by $s(\ell_B) \sim \ell_B^{-d_k}$
  \cite{song05}.  Scaling theory defines precise relations between the
  exponents for fractal networks \cite{song05}, through
  $\gamma=1+d_B/d_k$.  For the case of brain modules analyzed here
  (Fig.~\ref{fractal}E), we find $d_k=1.5\pm0.1$.  Using the values of
  $d_B$ and $d_k$ for the brain clusters, the prediction is
  $\gamma=2.26\pm0.11$, which is close to the calculated value of
  $\gamma=2.11\pm 0.04$ from Fig. \ref{fractal}D.

  The previous analysis reveals the mechanism of formation of a
  scale-free network, but it does not assure a fractal
  topology. Fractality can be determined from the study of the
  degree-degree correlation through the distribution, $P(k_1,k_2)$ to
  find a link between nodes with $(k_1, k_2)$ degree.  This
  correlation characterizes the hub-hub repulsion through scaling
  exponents $d_e$ and $\epsilon$ (see SI Appendix and
  Fig.~S5) \cite{song06,GallosPRL}.  In a fractal,
  they satisfy $\epsilon=2+d_e/d_k$.  A direct measurement of these
  exponents yields $d_e=0.51\pm0.08$ and $\epsilon=2.1\pm0.1$
  (Fig.~S5). Using the measured values of $d_e$ and
  $d_k$, we predict $\epsilon = 2.3\pm 0.1$, which is close to the
  observed exponent. Taken together, these results indicate a
  scale-free fractal morphology of brain modules. Such structure is in
  agreement with previous results of the anatomical connectivity of
  the brain \cite{meunier,bassett} and functional brain networks
  \cite{eguiluz2005scale}.

\subsection{ Quantifying submodular structure of brain modules}

Standard modularity decomposition methods \cite{newman1,fortunato}
based on maximization of the modularity factor $Q$ as defined in
\cite{newman1,fortunato,meunier,lazaros,galvao} are particularly
suitable to uncover the submodular structure. For example, the
  Girvan-Newman method \cite{newman1} yields a value of $Q\sim0.82$
  for the brain clusters, indicating a strong modular substructure.
The box covering algorithm benefits from detecting submodules (the
boxes) at different scales. Then, we can study the hierarchical
character of modularity \cite{meunier,lazaros,galvao}, and detect
whether modularity is a feature of the network that remains
scale-invariant (see SI Appendix and Fig.~S6 for
  a comparison of the submodular structure obtained using
  Girvan-Newman and box covering).

\begin{figure}
\centerline{ \resizebox{13.0cm}{!} { \includegraphics{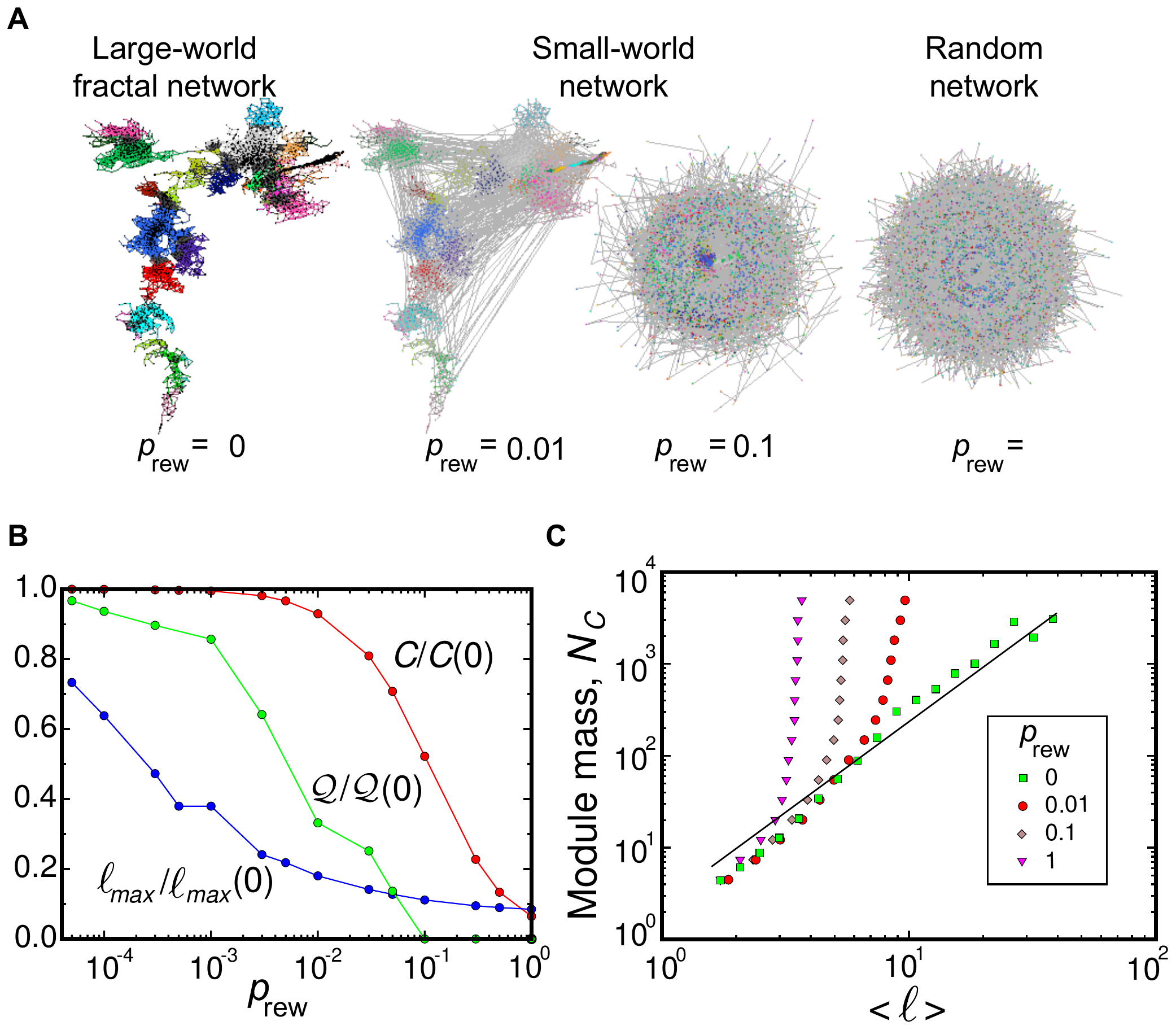}}}
 \caption{\label{box} {\bf Transition from fractal to small-world
     networks.}  {\bf (A)} Left panel shows a typical percolation
   module in network space. 
%MARIANO (subject \#11, SOA=0 in our dataset  \cite{dataset}). 
The colors identify sub-modules obtained by the
   box-covering algorithm with $\ell_B=15$.  This fractal module
   contains 4097 nodes with $\langle \ell \rangle = 41.7$, $\ell_{\rm
     max}=139$, and $r_{\rm max}=136$ mm.  When a small fraction
   $p_{\rm rew}$ of the links are randomly rewired \cite{watts98}, the
   modular structure disappears together with the shrinking path
   length.  The rewiring method starts by selecting a random link and
   cutting one of its edges. This edge is then rewired to another
   randomly selected node, and another random link starting from this
   node is selected. This is again cut and rewired to a new random
   node, and we repeat the process until we have rewired a fraction
   $p_{\rm rew}$ of links. The final link is then attached to the
   initially cut node, so that the degree of each node remains
   unchanged.  {\bf (B)} Small-world cannot coexist with modularity.
   The large diameter and modularity factor, Eq. (\ref{modular}) for
   $\ell_B=15$, of the fractal module in {\bf (A)} (left panel)
   diminish rapidly upon rewiring a tiny fraction $p_{\rm rew}\approx
   0.01$ of links, while the clustering coefficient still remains
   quite large.  {\bf (C)} The transition from fractal to small-world
   to random structure is shown when we plot the mass versus the
   average distance for all modules for different $p_{\rm rew}$ values
   as indicated.  The crossover from power-law fractal to exponential
   small-world/random is shown.  }
\end{figure}

The minimization of $N_B$ guarantees a network partition with the
largest number of intramodule links and the fewest intermodule
links. Therefore, the box covering algorithm maximizes the following
modularity factor \cite{lazaros,galvao}:
\begin{equation}
{\cal Q}(\ell_B) \equiv \frac{1}{N_B} \sum_{i=1}^{N_B} \frac{L_i^{\rm
    in}}{L_i^{\rm out}},
\label{mo}
\end{equation}
which is a variation of the modularity factor, $Q$, defined in
\cite{newman1,fortunato}.  Here, $L_i^{\rm in}$ and $L_i^{\rm out}$
represent the intra and intermodular links in a submodule $i$,
respectively.
Large values of $\cal Q$ (i.e. $L_i^{\rm out}\to 0$) correspond to
high modularity \cite{lazaros}. We make the whole modularization
method available at \cite{dataset}.

Figure \ref{fractal}F shows the scaling of ${\cal Q}(\ell_B)$ averaged
over all modules at percolation revealing a monotonic increase with a
lack of a characteristic value of $\ell_B$.  Indeed, the data can be
fitted with a power-law form \cite{lazaros}:
\begin{equation}
{\cal Q}(\ell_B) \sim \ell_B^{d_M} ,
\label{modular}
\end{equation}
which is detected through the modularity exponent, $d_M$.  We study the
networks for all the subjects and stimuli and find $d_M = 1.9\pm0.1$
(Fig. \ref{fractal}F).  The lack of a characteristic length-scale
expressed in Eq. (\ref{modular}) implies that
submodules are organized within larger modules
such that the inter-connections between those submodules repeat the
basic modular character of the entire brain network.

The value of $d_M$ reveals a considerable modularity in the system as
it is visually apparent in the sample of Fig. \ref{box}A, left panel,
where different colors identify the submodules of size $\ell_B=15$ in
a typical fractal module.  For comparison, a randomly rewired network
(Fig. \ref{box}A, right and central panels) shows no modularity and
has $d_M\approx 0$.  Scaling analysis indicates that $d_M$ is related
to $L_{\rm out}\sim \ell_B^{d_x}$, which defines the outbound exponent
$d_x$ characterizing the number of intermodular links for a submodule
\cite{lazaros} ($d_x$ is related to the Rent exponent in integrated
circuits \cite{bassett}).  From Eq. (\ref{modular}), we find: $d_M=
d_B - d_x$, which indicates that the strongest possible modular
structure has $d_M=d_B$ ($d_x=0$) \cite{lazaros}. Such a high
modularity induces very slow diffusive processes (subdiffusion) for a
random walk in the network \cite{lazaros}. Comparing
Eq. (\ref{modular}) with (\ref{db}), we find $d_x=0$, which quantifies
the maximum degree of modularity in the brain modules.

\subsection{Small-world or large-world fractal modularity}

An important consequence of Eqs. (\ref{df}) and (\ref{db}) is that the
network determined by the strong links above the first $p_c$-jump
lacks the logarithmic scaling characteristic of small-worlds
and random networks
\cite{watts98}:
\begin{equation}
\langle \ell \rangle \sim \log N_c,
\label{sw1}
\end{equation}
A fractal network poses much larger distances than those appearing in
small-worlds \cite{song05}: a distance $ \ell_{\rm max} \sim 100$
observed in Fig. \ref{fractal}A (red curve) would require an enormous
small-world network $N_c \sim 10^{100}$, rather than $N_c \sim
10^{4}$, as observed for fractal networks in Fig. \ref{fractal}A. The
structural differences between a modular fractal network and a
small-world (and a random network) are starkly revealed in the
panels of Fig. \ref{box}A. We rewire the fractal module on the left
panel by randomly reconnecting a fraction $p_{\rm rew}$ of the links
while keeping the degree of each node intact \cite{watts98}.

\begin{figure}
\centerline{ \resizebox{14.0cm}{!} {  \includegraphics{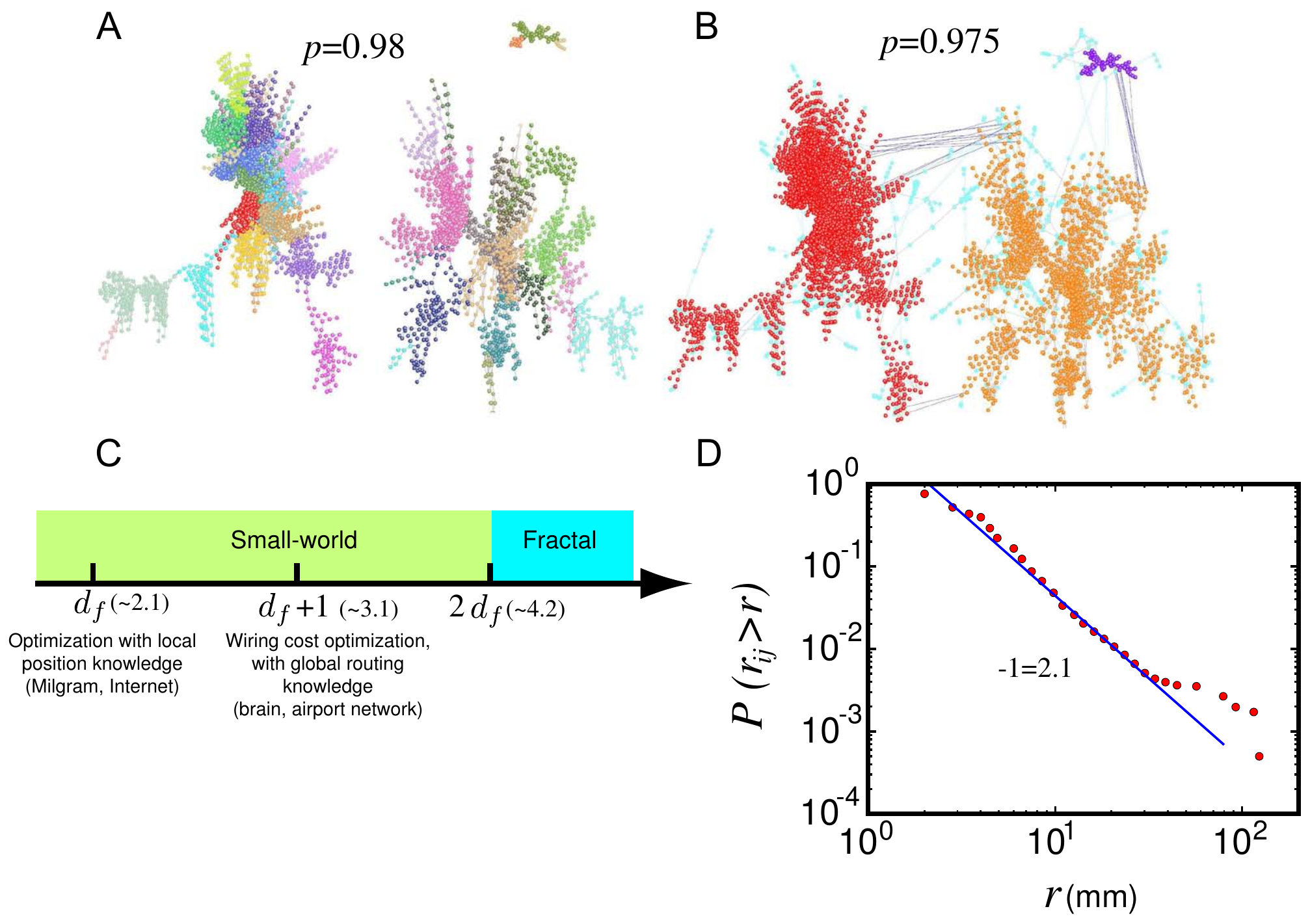}}}
\caption{\label{optimal} 
{\bf Weak ties are optimally distributed}.
{\bf (A)}
Three modules identified at $p_c=0.98$ for the subject in
Fig. \ref{perco}B. The colors correspond to different submodules as
identified by the box covering algorithm at $\ell_B=21$.
{\bf (B)}
When we lower the threshold to $p=0.975$, weak ties connect the
modules.  The three original modules as they appear in (A) are
plotted in red, orange and purple and the light blue nodes are the
nodes added from (A) as we lower $p$. Blue lines represent the
added weak links with distance longer than 10 mm. The weak links
collapse the three modules into one.
{\bf (C)}
Sketch of the different critical values of the shortcut exponent $\alpha$ in
comparison with $d_f$.
{\bf (D)}
Cumulative probability distribution
$P(r_{ij}>r)$. The straight line fitting yields an exponent
$\alpha-1=2.1\pm0.1$ indicating optimal information transfer with
wiring cost minimization \cite{shlomo}.  Certain clusters occupy
two diametric parts of the brain. In practice, these are two modules
that are connected through long-range links. These links increase
significantly the percentage of links at large distances $r_{ij}$,
since they are superimposed on top of the regular distribution of
links within unfragmented clusters. This behavior is manifested as a
bump in the curve.
}
\end{figure}

Figure \ref{box}B quantifies the transition from fractal ($p_{\rm
  rev}=0$) to small-world ($p_{\rm rev}\approx 0.01 - 0.1$) and
eventually to random networks ($p_{\rm rev}=1$), illustrated in
Fig. \ref{box}A: we plot $\ell_{\rm max}(p_{\rm rew})/\ell_{\rm
  max}(0)$, the clustering coefficient $C(p_{\rm rew})/C(0)$ and
${\cal Q}(p_{\rm rew})/{\cal Q}(0)$ for a typical $\ell_B=15$ as we
rewire $p_{\rm rew}$ links in the network.  As we create a tiny
fraction $p_{\rm rew}= 0.01$ of short-cuts, the topology turns
into a collapsed network with no trace of modularity left, while $C(0.01)$
still remains quite high (Fig. \ref{box}B).
The rewired networks present the exponential behavior of small-worlds
\cite{watts98}, and also random networks as $p_{\rm rev}$
  increases, obtained from Eq. (\ref{sw1}):
\begin{equation}
N_c \sim \exp\big({\langle \ell
  \rangle/\ell_0}\big),
\label{small}
\end{equation}
where $N_c$ is averaged over all the modules (Fig. \ref{box}C).  The
characteristic size is very small and progressively shrinks to
$\ell_0=1/7$ when $p_{\rm rew}=1$. The hallmark of small-worlds 
and random networks, exponential scaling Eq. (\ref{small}), is
incompatible with the hallmark of fractal large-worlds, power-law
scaling Eq. (\ref{db}).  Similarly, while we find a broad domain where
short network distances coexist with high clustering forming a
small-world behavior, modularity does not show such a robust behavior
to the addition of shortcuts.

\subsection{Short-cut wiring is optimal for efficient flow}

Figure \ref{box}B suggests that modularity and small-world cannot
coexist at the same level of connectivity strength. Next, we set out
to investigate how the small-world emerges. 

When we extend the percolation analysis lowering further the threshold
$p$ below $p_c$, weaker ties are incorporated to the network
connecting the self-similar modules through short-cuts. A typical
scenario is depicted in Fig. \ref{optimal}A showing the three largest
percolation modules identified just before the first percolation jump
in the subject \#1 shown in Fig. \ref{perco}B at $p=0.98$. For this
connectivity strength, the modules are separated and show submodular
fractal structure indicated in the colored boxes obtained with box
covering. When we lower the threshold to $p=0.975$,
Fig. \ref{optimal}B, the modules are now connected and a global
incipient component
starts to appears. A second global percolation-like transition appears
in the system when the mass of the largest component occupies half of
the activated area (see e.g. Fig. \ref{perco}).
For different individuals, global percolation occurs in the interval
$p=[ 0.945,0.96]$ as indicated in the inset of Fig. \ref{perco}A.

Our goal is to investigate whether the weak links shortcut the network
in an optimal manner.
When the cumulative probability distribution to find a Euclidean
distance between two connected nodes, $r_{ij}$, larger than $r$
follows a power-law:
\begin{equation}
  P(r_{ij}>r) \sim r^{-\alpha+1},
\label{alpha}
\end{equation}
statistical physics makes precise predictions about optimization
schemes for global function as a function of the short-cut exponent
$\alpha$ and $d_f$ \cite{kleinberg00,shlomo,rozenfeld}. Specifically,
there are three critical values for $\alpha$ as shown schematically in
Fig. \ref{optimal}C. If $\alpha$ is too large then shortcuts will not
be sufficiently long and the network will behave as fractal, equal to
the underlying structure. Below a critical value determined by $\alpha
< 2 d_f$ \cite{rozenfeld}, shortcuts are sufficient to convert the
network in a small world. Within this regime there are two significant
optimization values:

{\it (i)} Wiring cost minimization with full routing information.
This considers a network of dimension $d_f$, over which short-cuts are
added to optimize communication, with a wiring cost constraint
proportional to the total shortcut length.  It is also assumed that
coordinates of the network are known, i.e. it is the shortest path
that it is being minimized. Under these circumstances, the optimal
distribution of shortcuts is $\alpha = d_f + 1$ \cite{shlomo}.  This
precise scaling is found in the US airport network \cite{bianconi}
where a cost limitation applies to maximize profits.
		
{\it (ii)} Decentralized Greedy searches with only local information.
This corresponds to the classic Milgram's ``small-world experiment''
of decentralized search in social networks \cite{kleinberg00}, where a
person has knowledge of local links and of the final destination but
not of the intermediate routes.  Under these circumstances, which also
apply to routing packets in the Internet, the problem corresponds to a
greedy search, rather than to optimization of the minimal path. The
optimal relation for greedy routing is $\alpha=d_f$
\cite{kleinberg00,rozenfeld}.

Hence, the analysis of $P(r_{ij}>r)$ provides information both on the
topology of the resulting network and on which transport procedure is
optimized. This distribution reveals power-law behavior
Eq. (\ref{alpha}) with $\alpha=3.1\pm0.1$ when averaged over the
modules below $p_c$ (Fig. \ref{optimal}D).  Given the value obtained
in Eq. (\ref{df}), $d_f = 2.1$, this implies that the network composed
of strong and weak links is small-world ($\alpha<2d_f$)
\cite{rozenfeld} and optimizes wiring cost with full knowledge of
routing information ($\alpha=d_f+1$) \cite{shlomo}.

\section{Discussion}

The existence of modular organization which become small-world when
short-cut by weaker ties is reminiscent of the structure found to bind
dissimilar communities in social networks. Granovetter's work in
social sciences \cite{granovetter,kertesz} proposes the existence of
weak ties to cohese well-defined social groups into a large-scale
social structure.
The observation of such an organization in brain networks suggests that it may be a ubiquitous natural solution to the puzzle of information flow in highly modular structures.

Over the last decades, wire length minimization arguments have been
used successfully to explain the architectural organization of brain
circuitry
\cite{cowey1979cortical,linsker1986basic,mitchison1991neuronal,cherniak1995neural,chklovskii2000optimal}.
Minimizing wire length is in fact of paramount importance, since about
60\% of the cortical volume is taken up by wire (axons and dendrites)
\cite{chklovskii2002wiring}.  This turns out to optimize conduction
rate, posing a strict packing limitation of the amount of wire in
cortical circuits \cite{chklovskii2002wiring}.  Our finding of a
distribution of weak links which minimizes wiring cost is hence in
line with a previous literature, consistently showing that neural
circuit design is under pressure to minimize wiring length. However,
some important nuances of the specific optimization procedure ought to
be considered.  First, we specifically showed that at the mesoscopic
scale, short-cut distribution optimizes wiring cost while maintaining
network proximity.  This is consistent with the organization of
large-scale neural networks in which total wiring can in fact be
decreased by about 32\% (in 95 primate cortical areas) and up to 48\%
in the global neuronal network of the nematode Caenorhabditis elegans
\cite{kaiser2006nonoptimal}.  This extra wiring cost comes from
long-range connections which achieve network benefits of shortening
the distance between pairs of nodes \cite{kaiser2006nonoptimal}.  

Our results are in agreement with this observation, suggesting that
simultaneous optimization of network properties and wiring cost might
be a relevant principle of brain architecture. In simple words, this
topology does not minimize the total wire per-se, simply to connect
all the nodes; instead it minimizes the amount of wire required to
achieve the goal of shrinking the network to a small-world. A second
intriguing aspect of our results, which is not usually highlighted, is
that this minimization assumes that broadcasting and routing
information are known to each node.  How this may be achieved-- what
aspects of the neural code convey its own routing information--
remains an open question in Neuroscience.

  BOLD fMRI is an indirect measure of brain activity which relies
  on multiple vascular and biophysical factors which couple the neural
  response to the haemodynamic signal \cite{logothetis2001neurophysiological}. Even if in
  fMRI research it is always assumed that haemodynamic signals reflect
  metabolic demand generated by local neuronal activity, recent
  studies have shown reliable haemodynamic signals that entrains to
  task structure independently of standard neural predictors of
  haemodynamics \cite{sirotin2009anticipatory}.  Hence, our results, as any
  other fMRI analysis, have to be taken cautiously and may partly
  reflect the underlying structure of vascular motives. Specifically,
  the human cortical vascular system has a large number of arterial
  anastomoses which show a seemingly looking fractal structure in the
  mm to cm range \cite{duvernoy1981cortical}. Precise measurements of
  fractality have been reported at the micrometer scales in volumes of
  the order of a few mm$^3$ \cite{cassot2006novel,risser2006homogeneous}, which
  corresponds to approximately a voxel volume, where branching
  structure of microcapilarities then generates fractals. Hence, it is
  possible that the fractal organization of brain modules is inherited from the vascular system itself.

  Although we cannot readily test the influence of the vascular system
  at a large scale, it is still possible to address this concern
  at a microscopic scale, by discarding neighboring
  correlations. Neighboring voxels are expected to carry some shared signal due to
  spatial autocorrelations from the microvascular network.  To assure
  that our results do not rely on neighbouring correlations which
  might be particularly spurious, we coarse-grained the original fMRI
  signal by doubling the lattice spacing, reducing the number of
  voxels by a factor of 8 and repeat the calculations.  The results
  are consistent with the percolation picture of fractal modules,
  albeit with an expected lower $p_c$. Such a renormalized $p_c$ is
  expected from renormalization theory to change under
  coarse-graining, while the main results on long-range links, such as
  the value of the exponents, are insensitive to this type of
  coarse-graining.  

  We also investigate whether the map of fractal dimension $d_B$
  reflects a meaningful organization based on known facts of
  functional properties of the cortex and the specific task which
  subjects are performing.  We found a topographical organization of
  fractality in the human brain (Fig.~S7).  The right
  portion of the anterior cingulate, SMA and the right PPC regions
  involved in routing of information and cognitive control
  \cite{zylberberg2011human,duncan2010multiple}, which are expected to
  have a more complex functional organization, are the clusters with
  higher fractal dimension. The left-right asymmetry is interesting
  since, in this specific task, the left hand response is queued for a
  few hundred millisencods and has to be temporally connected to
  working memory and inhibitory circuits.  While not fully conclusive,
  this analysis suggests a functional role of the network
  architectures described here.

  Another similar concern is that the recovered brain modules may 
  be a manifestation of the fractal structure of the underlying
  three-dimensional vortex grid or of the cortex. However, since the
  dimensions of the grid ($d=3$) and of the cortex
  ($d=2.7$)\cite{kiselev} are both sufficiently different from 1.9 and
  the connectivity distribution of the modules is much broader than
  the typical Euclidean fractal cortex (which should be narrow around
  $k\sim6$) or a 3d lattice ($k=6$), we may safely assume that these
  objects have their own structure. Moreover, we also observed modules
  with similar fractal dimension in subcortical structures suggesting
  that these results do not simply reflect anatomical properties of
  the cortical mantle.

A hierarchical modular organization of the brain composed of modules
within modules has been invoked in \cite{meunier,bassett} to describe
the brain structure.
The present results support these previous findings, while, at the
same time, provide a new view by integrating the results with the
(non-critical) properties of small-worlds and placing self-similarity
in the framework of
scaling theory, universality and Renormalization Groups
\cite{stanley2}.  In this framework, brain modules are characterized
by a set of novel scaling exponents, the septuplet: $(d_f, d_B, d_k,
d_e, d_M,\gamma, \alpha) = (2.1, 1.9, 1.5, 0.5, 1.9, 2.1, 3.1)$, and
the scaling relations $d_M=d_B-d_x$, relating fractality with
modularity, $\alpha=d_f+1$, relating global integration with
modularity, $\gamma=1+d_B/d_k$, relating scale-free with fractality,
and $\epsilon = 2 + d_e/d_k$, relating degree correlations with
fractality.  

One advantage of this formalism is that the different brain topologies
can be classified into universality classes under RG \cite{stanley2}
according to the septuplet $(d_f, d_B, d_k, d_e, d_M, \gamma,
\alpha)$.  Universality applies to the critical exponents but not to
quantities like $(p_c, C, \ell_0)$ which are sensitive to the
microscopic details of the different experimental situations
\cite{stanley2}. In this framework, (non-critical) small-worlds are
obtained in the limit $(d_f, d_B, d_k, d_e, d_M, d_x) \to (\infty,
\infty, \infty, 0, 0, \infty)$.  A path for future research will be to
test the universality of the septuplet of exponents
under different activities covering other areas of the brain, e.g.,
the resting-state correlation structure \cite{raichle}.

In conclusion, we propose a formal solution to the problem of
information transfer in the highly modular structure of the brain. The
answer is inspired by a classic finding in sociology: the strength of
weak ties \cite{granovetter}.  
The present work provides a general insight into the physical
mechanisms of network information processing at large. It builds up on
an example of considerable relevance to natural science, the
organization of the brain, to establish a concrete solution to a broad
problem in network science. The results can be readily applied to
other systems--- where the coexistence of modular specialization and
global integration is crucial--- ranging from metabolic, protein and
genetic networks to social networks and the Internet.

\begin{acknowledgments}
  LKG and HAM thank the NSF-0827508 Emerging Frontiers Program for
  financial support.  MS is supported by a Human Frontiers Science
  Program Fellowship.  We thank D. Bansal, S. Dehaene, S. Havlin, and
  H.D. Rozenfeld for valuable discussions.

\end{acknowledgments}

%% Add your bibliography items here.  PNAS requires that bibliography items
%% be entered directly into the article rather than called from a BibTeX
%% environment.  Contact pnas@nas.edu if you need assistance with your
%% bibliography.

% Sample bibliography item in PNAS format:
%% \bibitem{Ch} D. Chae (2003) {\it Nonlinearity} {\bf 16}, 479-495.
%% \bibitem{in-text reference} Author Names (year published)
%% {\it Journal Name} {\bf Volume #}, start page-end page

%\newpage 

\clearpage
\newpage

\setcounter{figure}{0}
\renewcommand{\thefigure}{S\arabic{figure}}

%\arabic{section}
\setcounter{section}{0}

\centerline{\bf \Large SUPPORTING INFORMATION}

\section{fMRI methods and network construction}
\label{fmri}

A total of 16 participants (7 women and 9 men, mean age, 23, ranging
from 20 to 28) were asked to perform two tasks with the instruction
that they had to respond accurately and fast to each of them. The
first task was a visual task of comparing a given number (target T1)
to a fixed reference, and, second, an auditory task of judging the
pitch of an auditory tone (target T2) [36]. The two
stimuli are presented with a stimulus onset asynchrony (SOA), i.e.,
the delay in the onset of T1 and T2, varying from: SOA=0, 300, 900 and
1200 ms.  In the number-comparison task, a number varying randomly
among four values (28, 37, 53, 62) was flashed on a computer screen
and subjects had to respond, with a key press using the right hand,
whether the number was larger or smaller than 45. In the auditory
task, subjects had to respond whether the tone was high or low
frequency with a key press using the left hand.  Full details and
preliminary statistical analysis of this experiment have been reported
in [36]. The study is part of a larger neuroimaging
research program headed by Denis Le Bihan and approved by the Comit\'e
Consultatif pour la Protection des Personnes dans la Recherche
Biom\'edicale, H\^{o}pital de Bic\^{e}tre (Le Kremlin-Bic\^{e}tre,
France).

Subjects performed a total of 160 trials (40 for each SOA value) with
a 12 s inter-trial interval
[37].  The 160 trials were performed in five
blocks of 384 s with a resting time of $\sim$ 5 min between blocks.
For each trial, we recorded whole-brain fMRI images at a sampling
time, TR = 1.5 s producing 8 fMRI images between two consecutive
trials.  From these images we computed the phase and amplitude of the
hemodynamic response of each trial as explained in
[37].  The experiments were performed on a 3T fMRI
system (Bruker).  Functional images sensitive to blood oxygenation
level dependent (BOLD) contrast were obtained with a T2$^*$-weighted
gradient echoplanar imaging sequence [repetition time (TR) = 1.5 s;
echo time = 40 ms; angle = 90$^{\circ}$; field of view (FOV) = 192
$\times$ 256 mm; matrix = 64 $\times$ 64]. The whole brain was
acquired in 24 slices with a slice thickness of 5 mm.  High-resolution
images (three-dimensional gradient echo inversion-recovery sequence,
inversion time = 700 mm; FOV = 192 $\times$ 256 $\times$ 256 mm;
matrix = 256 $\times$ 128 $\times$ 256; slice thickness = 1 mm) were
also acquired.

To estimate the periodicity and phase of the event-related BOLD
response, the data from each subject were submitted to a first-level
model in which the signal from each trial (8 TRs of 1.5 s) was fitted
with three regressors: a constant, a sine, and a cosine function at
the above period. To facilitate intersubject averaging across possible
differences in anatomical localization, the regression weights of the
sines and cosines were stereotactically transformed to the
standardized coordinate space of Talairach and Tournoux ([Montreal
Neurological Institute] MNI 152 average brain) to spatially normalize
for individual differences in brain morphology. Normalized images had
a resolution of 8 mm$^3$. Normalized phase images were transformed
with the inverse tangent function to yield a phase
lag expressed in radians
for each voxel $i$ and each trial $t=1, .. , T$ over $T=40$ trials:
$\phi_i(t) \in [0,2\pi]$ [37], indicating phase
lags in the interval $[0,12]$s.

We calculate cross-correlations between different brain areas based on
these phases [11, 12, 38].
% when a subject responds to the external stimulus.  
We determine the equal-time cross-correlation matrix $\mathbb{ C}$
with elements $C_{ij}$ measuring the cross-correlation between the
phase activity $\phi_i(t)$ of the $i$-th and $j$-th voxel over $T=40$
trials for each subject and SOA condition:

\begin{equation}
C_{ij} = \frac{1}{T} \sum_{t=1}^T \cos(\phi_i(t)-\phi_j(t)).
\label{c}
\end{equation}
By construction, the elements satisfy $-1 \le C_{ij} \le 1$, where
$C_{ij} =1$ corresponds to perfect correlations, $C_{ij} =-1$
corresponds to perfect anticorrelations, and $C_{ij} = 0$ describes a
pair of uncorrelated voxels.
The entire experimental dataset is available in [40].

For our analysis, we create a mask where we keep voxels which were
activated in more than 75\% of the cases, i.e., in at least 48
instances out of the 64 total cases considered.  The obtained number
of activated voxels is $N\approx 60,000$, varying slightly for
different individuals and stimuli. The `activated or functional map'
exhibits phases consistently falling within the expected response
latency for a task-induced activation [36]. As
expected for an experiment involving visual and auditory stimuli and
bi-manual responses, the responsive regions included bilateral visual
occipito-temporal cortices, bilateral auditory cortices, motor,
premotor and cerebellar cortices, and a large-scale bilateral
parieto-frontal structure, see SI-Section ``Spatial projection of the
modules'' below .  In the present analysis, we do not explore the
differences in networks between different conditions.  Rather, we
consider them as independent experiments, generating a total of 64
different networks, one for each condition of temporal gap and
subject.

The use of fMRI neighboring voxels can be expected to carry some
shared signal due to spatial autocorrelations (vascular, subject
motion or scanner noise), which could give rise to spurious
correlations over short distance. To test for this effect, we double
the lattice spacing, reducing the voxels by a factor of 8 and repeat
the calculations.  The results are consistent with the percolation
picture of Fig.~1, albeit with a lower $p_c$, while the main
results on long-range links are insensitive to this type of artifacts.

\begin{figure}
\centerline{ \resizebox{8.0cm}{!} { \includegraphics{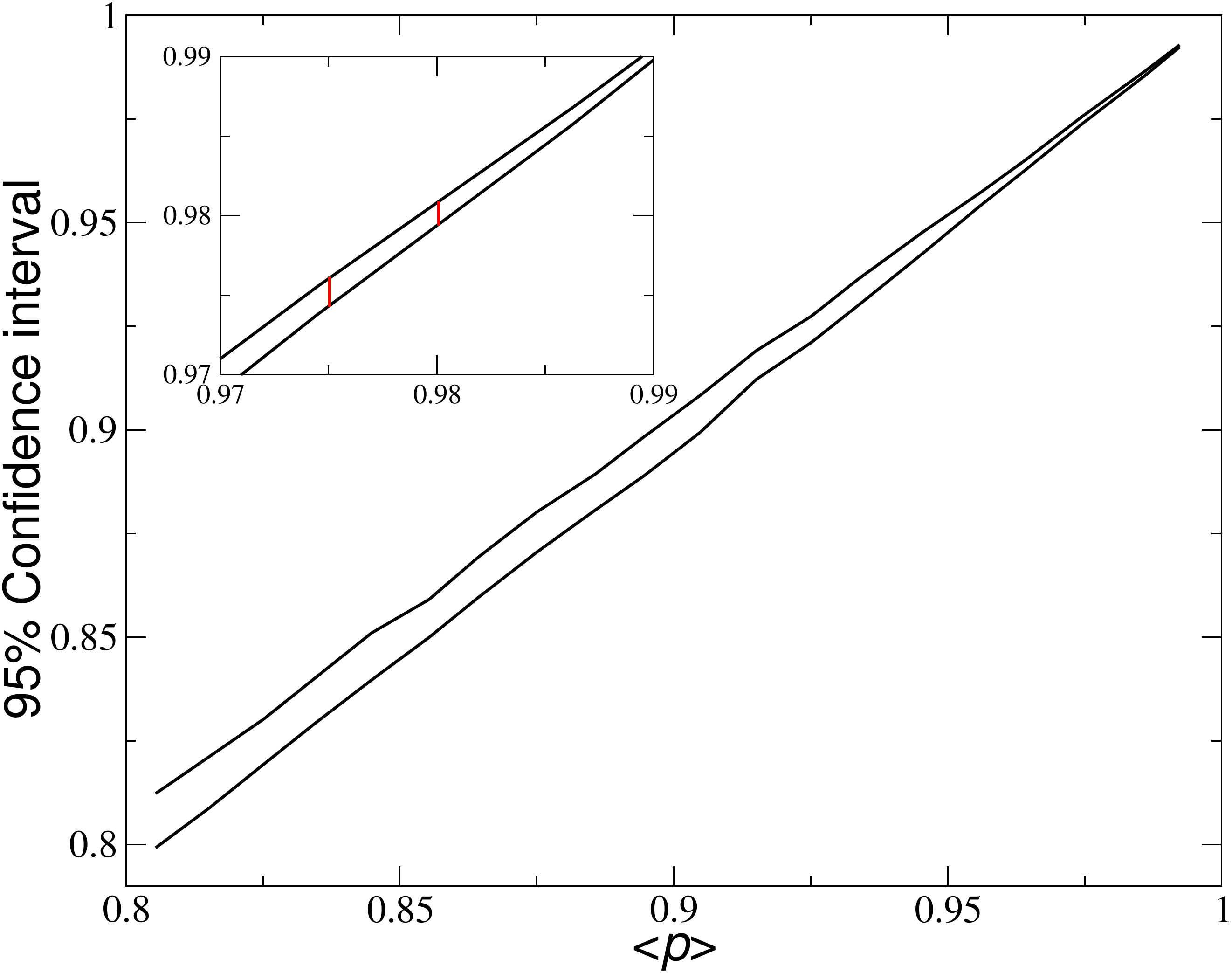}}}
  \caption{\label{si_bs}
{\bf Boot strap analysis}. The interval between the two curves corresponds to
the 95\% confidence interval for the calculation of the mean fraction
of links $\langle p \rangle$ as a function of $\langle p \rangle$. The
inset zooms in the regime around the values used in Fig. 4A.}
\end{figure}

\section{Boot strap analysis}

In order to estimate the accuracy of the correlation calculations, we
performed a nonparametric boot strap analysis. We consider the set of
the 40 trials per subject and SOA value. We perform the boot strap
analysis for each possible pair of voxels.  The correlation between
two voxels for each of those trials serves as our original sample of
40 correlation values. We then draw 10000 re-samples from this sample
with substitution. The arithmetic mean is calculated for each
re-sample.  Calculating the average value of all these means gives the
boot strap estimate for the mean correlation. The 95\% boot strap
confidence interval is calculated by the distribution of the 10000
mean values at the 0.05 and 0.95 points of the distribution,
respectively.

The above process yields the confidence interval for the correlation
value between two voxels. A different pair of voxels may have very
different value of correlation, so in Fig.~\ref{si_bs} we present
the 95\% bootstrap confidence interval as a function of the average
value of correlation. The interval becomes smaller, i.e. the accuracy
of the calculation increases, for larger $p$ values.  Considering the
networks of Fig.~4A and B, for example, the intervals for
$p=0.975$ and $p=0.98$ correspond to (see inset) [0.9744, 0.9760] and
[0.9795, 0.981], respectively.

\section{Spatial projection of the modules} 
\label{SpatialProjection} 

The complex network representation reveals functional links between
brain areas, but cannot directly reveal spatial correlations. Since
voxels are embedded in real space, we also study the topological
features of modules in three dimensions, where now voxels assume their
known positions in the brain and links between them are transferred
from the corresponding network, i.e., they are assigned according to
the degree of correlation between any two voxels, Eq. (\ref{c}), which
is independent of the voxels proximity in real space.  The above
procedure yields a different spatial projection of the modules for
each subject; an example for subject \#1 and SOA=900 ms in the medial
occipital cortex is shown in Fig.~1D.  We study each of
these percolation modules separately and find that they all carry
statistically similar patterns. The topography of the identified
modules reflects coherent patterns across different subjects, as shown
next.

Fig.~S2A shows a medial sagital view of the largest
four percolation modules for all the participants under stimulus
SOA=0.  In virtually all subjects we observe a module covering the
anterior cingluate (AC) region, a module covering the medial part of
the posterior parietal cortex (PPC) and a module covering the medial
part of posterior occipital cortex (area V1/V2), along the calcarine
fissure.

We measure the likelihood that a voxel appears in the largest
percolation module among all the participants in Fig.~S2A by
counting, for each voxel, the number of individuals for which it was
included in one of the first four percolation modules.  The spatial
distribution of the first percolation modules averaged over all the
subjects depicted in Figs.~S2B and S2C shows
that modules in the three main modes, V1/V2, AC and PPC, are
ubiquitously present in percolation modules and, to a lesser extent,
voxels in the motor cortex (along the central sulcus) are slightly
more predominantly on the left hemisphere.  The correlation networks
obtained from each subject yield modules with consistent topographic
projections.

\begin{figure}
\centerline{ \resizebox{8.0cm}{!} { \includegraphics{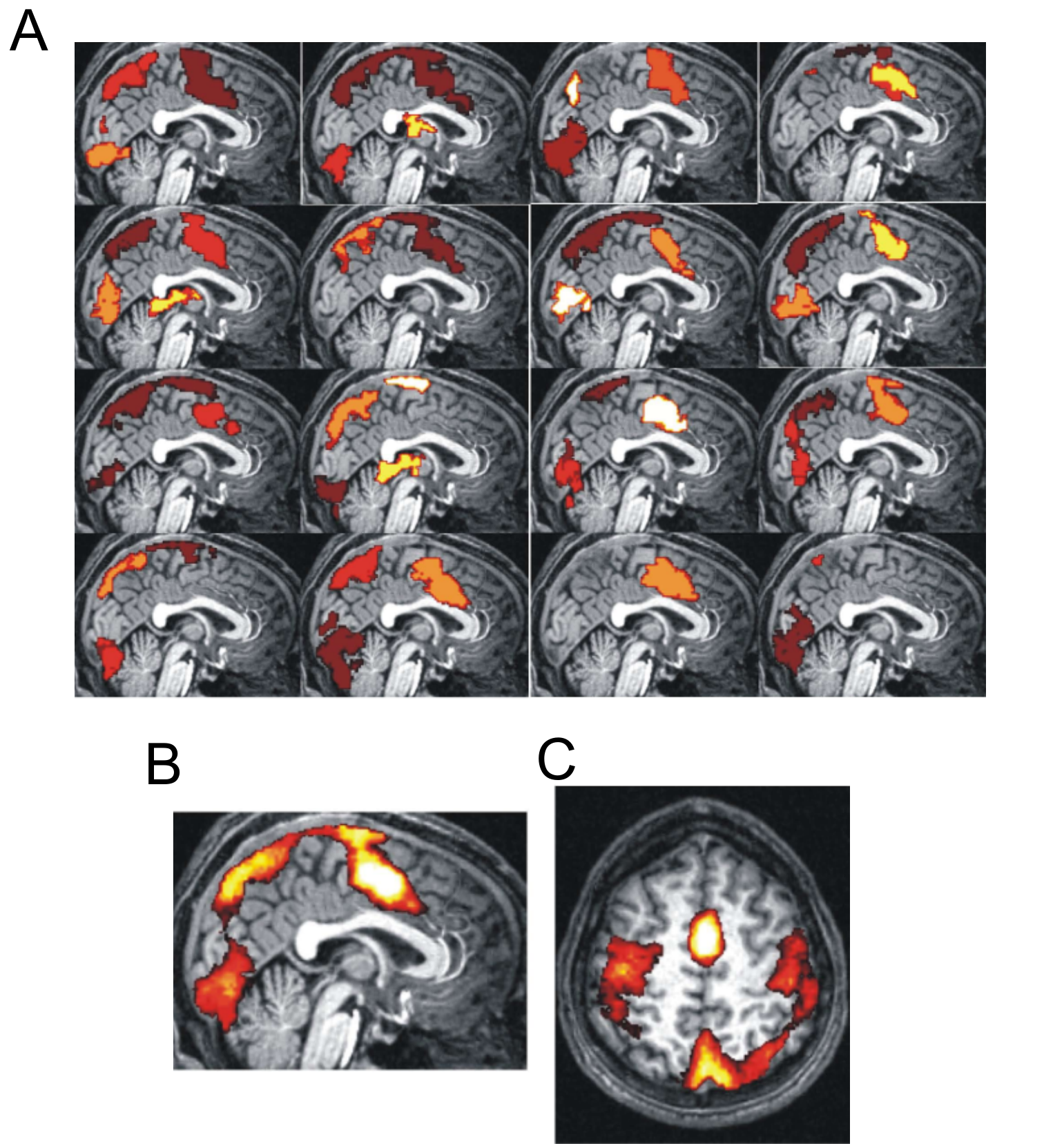}}}
\caption{\label{sagital}  {\bf The emerging modules have
    consistent spatial projections}.  {\bf (A)} Spatial distribution
  of the four largest percolation modules (yellow, orange, red, brown)
  appearing at the first percolation jump, $p_c$, for each subject
  under stimulus SOA=0. Most modules are localized in the same
  regions: anterior cingulate, posterior medial-occipital, posterior
  parietal and thalamus.  {\bf (B)} and {\bf (C)} These panels show
  the number of times that the largest percolation cluster for each of
  the 16 subjects appears in a given voxel.  White bleached regions
  correspond to voxels which are active in the 16 subjects, while the
  red regions correspond to voxels shared by half of the subjects. The
  anterior cingulate, a fundamental node in cognitive control, is the
  only region shared by all subjects.  }
\end{figure}

\section{Box covering algorithm for fractal dimension in
  network space}
\label{memb}

For a given percolation module, the detection of submodules or boxes
follows from the application of the box-covering algorithm for
self-similar networks [22,43]. The algorithm can be
downloaded at [40].  In box covering we assign every node to
a box or submodule, by finding the minimum possible number of boxes,
$N_B(\ell_B)$, that cover the network and whose diameter (defined as
the maximum distance between any two nodes in this box) is smaller
than $\ell_B$.

We implement the Maximum Excluded Mass Burning (MEMB) algorithm from
[43] for box covering.  The algorithm uses the basic idea of
box optimization, where we require that each box should cover the
maximum possible number of nodes, and works as follows: We first
locate the optimal `central' nodes which will act as the origins for
the boxes.  This is done by first calculating the number of nodes
(called the mass) within a distance $r_B$ from each node. We use,
$\ell_B=2 r_B + 1$.  The node that yields the largest mass is marked
as a center.  Then we mark all the nodes in the box of this center
node as `tagged'.  We repeat the process of calculating the mass of
the boxes starting from all non-center nodes, and we identify a second
center according to the largest remaining mass, while nodes in the
corresponding box are `tagged', and so on.  When all nodes are either
centers or `tagged' we have identified the minimum number of centers
that can cover the network at the given $r_B$ value. Starting from
these centers as box origins, we then simultaneously burn the boxes
from each origin until the entire network is covered, i.e. each node
is assigned to one box (we call this process burning since it is
similar to burning algorithms developed to investigate clustering
statistics in percolation theory [29,30]).  In
Fig.~2A we show how box-covering works for a simple network at
different $\ell_B$ values. RG is then the iterative application of
this covering at different $\ell_B$.

\section{Correlation function}

Connections between voxels are determined according to the value of
the correlation between the two voxels, as described above.  This
value may also depend on the physical (Euclidean) distance between the
two voxels, since areas that are close to each other should interact
stronger.

\begin{figure}
\centerline{ \resizebox{8.0cm}{!} { \includegraphics{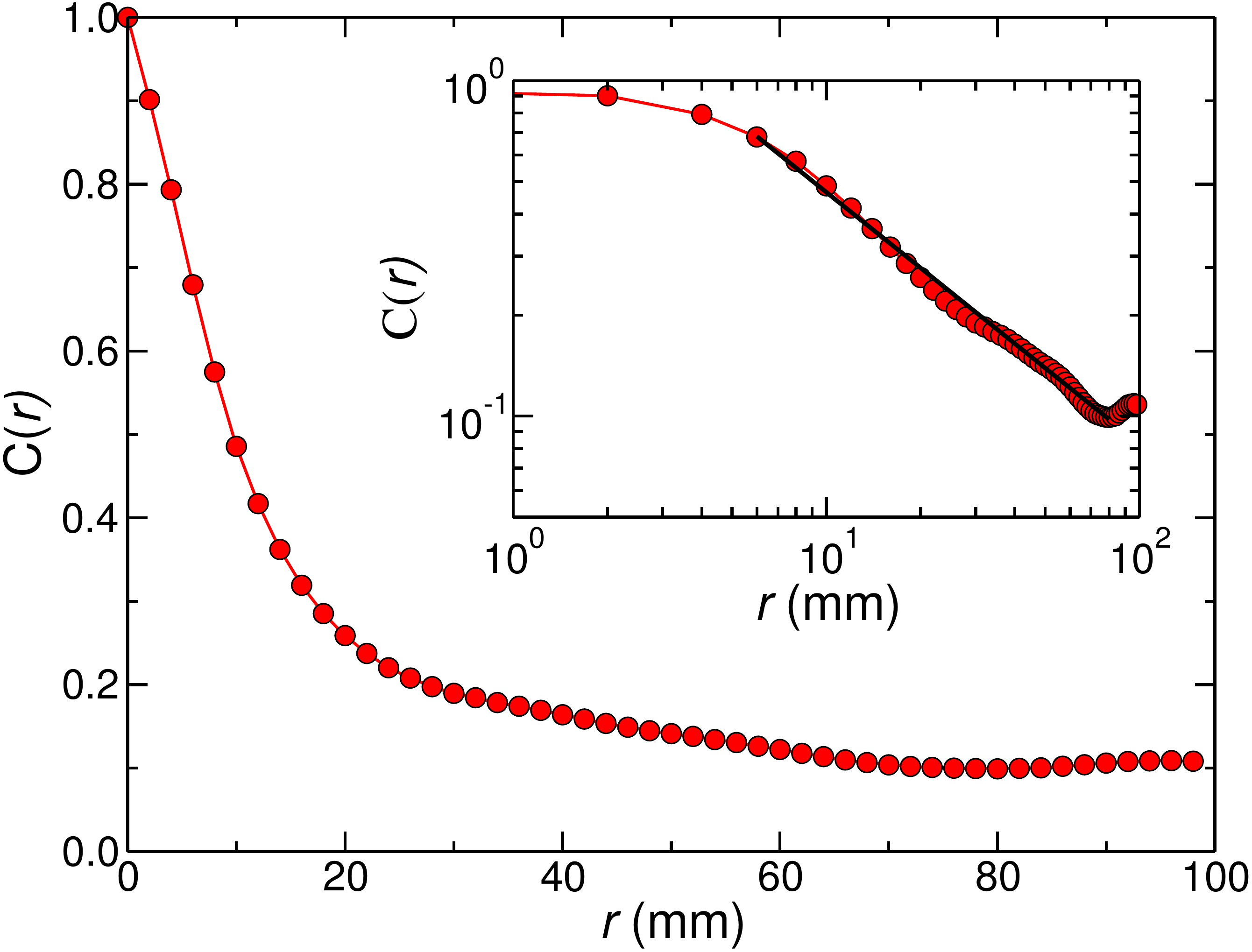}}}
\caption{\label{si_correl} {\bf Spatial correlation
    function}. This function measures the correlation $C(r)$ between
  the phase of two voxels that are at a Euclidean distance $r$ apart,
  as a function of $r$. As shown in the inset, it decays as a power law
  with slope $0.75\pm0.02$.}
\end{figure}

We studied the correlation function, $C(r)$ of the phases of the
voxels:
\begin{equation}
C(r) = \langle \cos(\phi_1-\phi_2) \rangle  ,
\end{equation}
where $\phi_i$ denotes the phase of voxel $i$. The distance $r$ is the
Euclidean distance between the two voxels 1 and 2 and the average is
taken over all pairs at distance $r$.  This function can be
interpreted as the correlation between two spins with orientation
determined by the phases $\phi_i$ of the voxels. We notice that this
correlation function is usually studied in Ising-like spin models. We
find that $C(r)$ decays algebraically with distance, as shown in
Fig.~\ref{si_correl}, and follows a power law form, $C(r)\sim r^{0.75}$.
The value of the exponent $0.75\pm0.02$ was calculated through standard
OLS regression. Notice that this function does not go to 0
asymptotically, but reaches a value of 0.1, which represents the
average correlation (notice that in the definition of the correlation,
the average value was not subtracted). This indicates that long-range
correlations remain strong even at large distances. Further
analysis is required to elaborate on this point, which is currently
outside the scope of our present study.

\section{Exponents calculation}

In Fig. 2D of the main text we show an aggregate average of the degree
distributions for all clusters. This curve exhibits the general trends
of the $P(k)$ distribution, demonstrating for example the heavy tail,
but it cannot be used for a direct determination of the exponent $\gamma$.

In our work we studied the properties of 192 network clusters, as described
in the main text. The calculation of the scaling exponents was done separately
for each network. The resulting set of 192 values was then analyzed through
non-parametric boot strap analysis, in order to get the average value of the
exponent and the corresponding confidence intervals.

\begin{figure}
\centerline{ \resizebox{12.0cm}{!} { \includegraphics{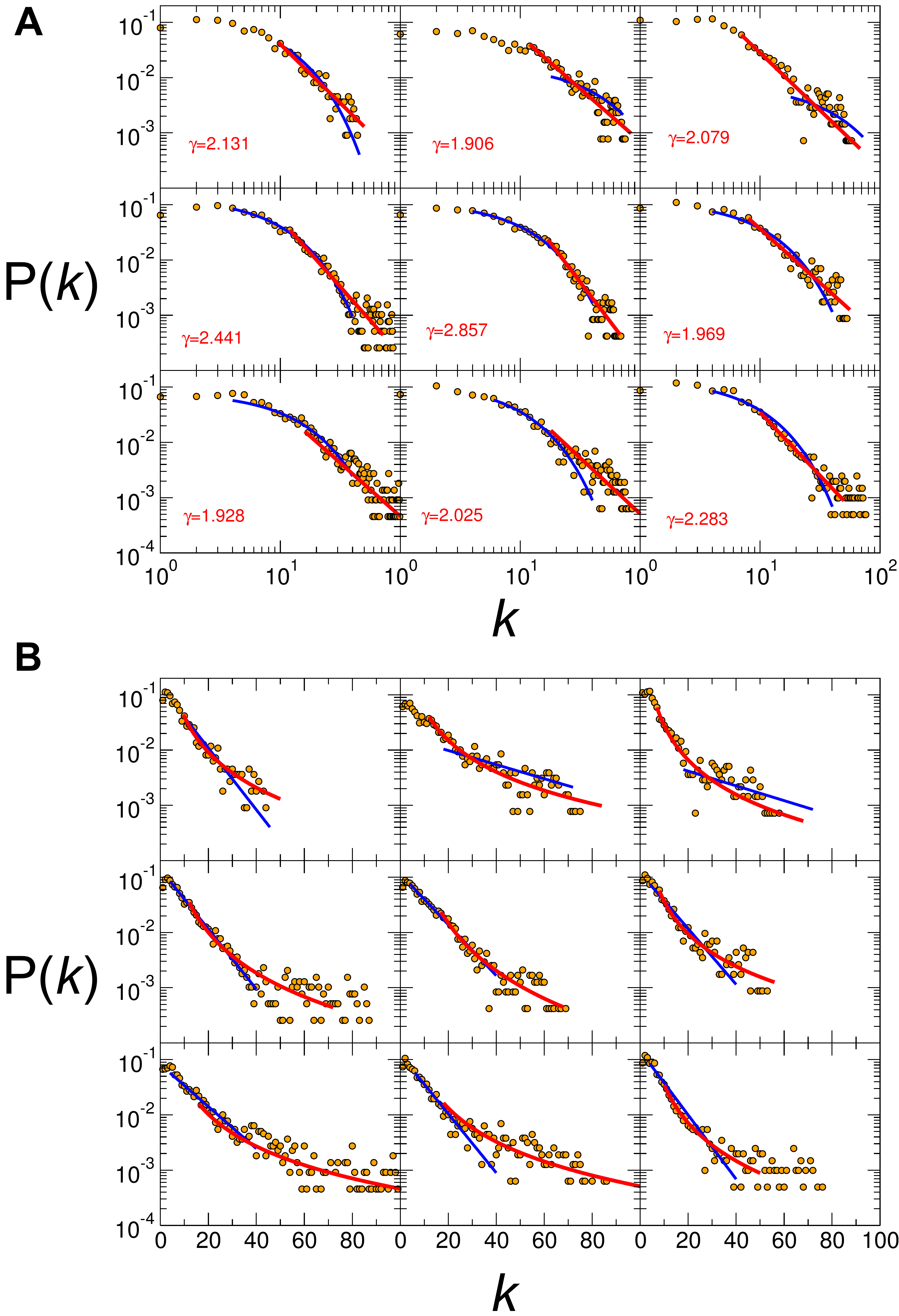}}}
\caption{\label{si_Pk} {\bf Degree distribution for network clusters}.
  A number of degree distribution functions $P(k)$ are shown for different
  clusters. The red lines correspond to the best power-law fitting, and
  the blue ones to an exponential fitting. {\bf (A)} Degree distribution $P(k)$
  in logarithmic axes. The power-law slopes correspond to the exponent $\gamma$,
  and are shown on the plots. {\bf (B)} The same distributions and fittings
  in semi-logarithmic axes.
  }
\end{figure}

As an example, in Fig.~\ref{si_Pk} we show the degree distributions for
9 different clusters. In the plots, it is clear that there is always a plateau
at small $k$ values, while in many cases there is an asymptotic exponential cutoff.
We fitted these distributions assuming that a power law describes the data within
a given interval only. For this, we used a generalized power-law form
\begin{equation}
P(k;k_{min},k_{max}) = \frac{k^{-\gamma}}{\zeta(\gamma,k_{min})-\zeta(\gamma,k_{max})} ,
\end{equation}
where $k_{min}$ and $k_{max}$ are the boundaries of the fitting interval
and the Hurwitz $\zeta$ function is given by $\zeta(\gamma,\alpha)= \sum_i (i+\alpha)^{-\gamma}$.

We used the maximum likelihood method, following e.g. Clauset et al, SIAM Review, 51, 661 (2009).
The fit was done in an interval where the lower boundary was $k_{min}$.
For a given $k_{min}$ value we were fixing the upper boundary to $k_{max}=w k_{min}$, where $w$
is a parameter. We calculated the slopes in successive intervals by continuously increasing $k_{min}$
and varying the value of $w$ from 4 to 30. In this way, we sampled a large number of possible
intervals. For each one of them we calculated the maximum likelihood estimator through the numerical solution of
\begin{equation}
\gamma = \rm{argmax} \left( -\gamma \sum_{i=1}^N \ln k_i - N \ln \left[ \zeta(\gamma,k_{min})-\zeta(\gamma,k_{max}) \right] \right)
\end{equation}
where $k_i$ are all the degrees that fall within the fitting interval and $N$ is the total number of
nodes with degrees in this interval. The optimum interval was determined through the
Kolmogorov-Smirnov test.

For the goodness-of-fit test, we used the Monte-Carlo method described in Clauset et al.
For each possible fitting interval we generated 10000 synthetic random distributions following
the best-fit power law. We then calculated the value of the Kolmogorov-Smirnov (KS) test for
each one of them and measured the fraction $p_{fit}$ of realizations where the real data KS value
was smaller than the synthetic SK value. We accepted the power-law hypothesis when this
ratio was larger than $p_{fit}>0.2$. The average ratio over all clusters that were retained was $p_{fit}=0.65$.
In this way, it is possible that we could
accept more than one exponents for a given cluster at different intervals.
In all these cases, the different $\gamma$ values were very close to each other and we considered
the final exponent to be the average of the individual exponents.

Standard boot strap analysis on the resulting set of the individual cluster values yielded the exponent
$\gamma=2.11\pm0.04$, with a 95\% confidence interval $[2.039,2.178]$.

The same analysis was performed to test for a possible exponential description of the data. We scanned
the same intervals as for the case of power-law and we used the maximum likelihood method to determine
the optimum exponential fitting to the form:
\begin{equation}
P(k;k_{min},k_{max}) = \frac{1-e^{-\lambda}}{e^{-\lambda k_{min}}- e^{-\lambda k_{max}}} e^{-\lambda k} .
\end{equation}
We again used KS statistics to determine the optimum fitting intervals and also the goodness-of-fit.
In all the cases where the power-law was accepted, the exponential fitting gave an average ratio
of $p_{fit}=0.017$, which rules out the possibility of an exponential distribution.

\section{Scaling analysis}

The structure of a fractal network can be characterized by a set of
scaling exponents.  They define the scaling of many important system
properties. Some of these properties and the corresponding exponents
are as follows:

a. The degree distribution: $P(k)\sim k^{-\gamma}$, where $\gamma$ is
the degree exponent [44].

b. The scaling of the mass with size: $N_B\sim \ell_B^{-d_B}$, which
defines the fractal exponent $d_B$ [22].

c. The degree-degree distribution $P(k_1,k_2) \sim k_1^{-\gamma+1}
k_2^{-\epsilon}$, where $\epsilon$ is the degree-degree exponent, and
can be measured through $E_b(k)\sim k^\epsilon$, which is the
integration of $P(k_1,k_2)$ over $k_2$ [48].

d. The probability that modules are connected through their hubs,
${\cal E} \sim \ell_B^{-d_e}$ defines the hub-hub exponent $d_e$
[23].

e. The scaling of the degree of the modules with the size of the
modules: $s \sim \ell_B^{-d_k}$, which defines the $d_k$ exponent
[22].

f. The scaling of the modular factor as defined in Eq.~(3):
${\cal Q}(\ell_B) \sim \ell_B^{d_M}$, through the modularity exponent
$d_M$ [27,28].

Scaling theory then defines precise relations between the exponents
valid for fractal scale-free networks:

g. $\gamma=1+d_B/d_k$ [22], 

h. $\epsilon=2+d_e/d_k$ [48], and 

i. $d_M = d_B - d_x$ [27,28].

\begin{figure}
\centerline{ \resizebox{16.0cm}{!} { \includegraphics{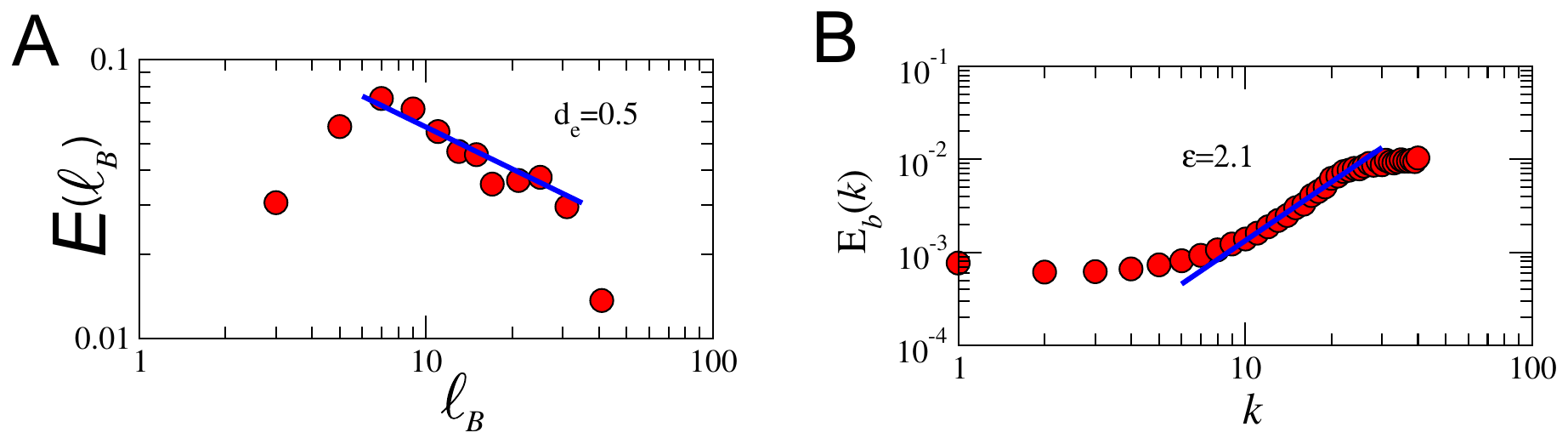}}}
  \caption{\label{si_matrix}
{\bf Calculation of the scaling exponents}.  
%{\bf (A)} Calculation 
%of the degree exponent $\gamma$ through the degree distribution $P(k)$.
%{\bf (B)} Fractal dimension $d_B$ through the dependence of the number
%of boxes of size $\ell$ as a function of $\ell$. {\bf (C)} Exponent $d_k$
%through the scaling of $s(\ell_B)$ vs $\ell_B$. 
{\bf (A)} Hub-hub exponent $d_e$ through the scaling of ${\cal
  E}(\ell_B)$. {\bf B} Degree-degree exponent $\epsilon$ through the
dependence of $E_b(k)$ on the degree $k$ [48].}
\end{figure}

We have measured directly all the exponents (see Fig.~2
and Fig.~\ref{si_matrix}) for the brain modules and find:
$\gamma=2.11\pm0.04$, $d_e=0.51\pm0.08$, $d_B=1.9\pm0.1$, $d_k=1.5\pm0.1$,
$\epsilon=2.1\pm0.1$, $d_M=1.9\pm0.1$.
Using these values in the known scaling relations above
(g) and (h), we predict $\gamma=2.26\pm0.11$ and $\epsilon=2.34\pm0.06$, which are
reasonably
close to the calculated exponents $\gamma=2.11$ and $\epsilon=2.1$
from the direct measures. This set of results gives support to a
scale-free fractal morphology of the brain modules. Notice that a
Euclidean 2d lattice would be obtained in the limit $\gamma\to\infty$,
$d_k$=0, $\epsilon\to\infty$.

\section{Modularity analysis}

In the main text of the paper we have described our modularity
  analysis of the brain clusters according to the MEMB technique. The
  modular properties of the same clusters can be also analyzed through
  techniques that partition a network according to maximization of
  modularity. We employed the Girvan-Newman method [20],
  which locates the point where the modularity measure, $Q$, is
  maximum.  The definition of $Q$ according to [20] is:
\begin{equation}
  Q = \sum_{i=1}^{N_M} \left( \frac{l_i}{L} - \left( \frac{d_i}{2L} \right)^2 \right) ,
\end{equation}
where $N_M$ is the number of modules, $L$ is the number of links in
the network, $l_i$ is the number of links within the module $i$, and
$d_i$ is the sum of the degrees in this module. A value of $Q=0$
corresponds to a completely random configuration or to the case of one
module only.

\begin{figure}
\centerline{ \resizebox{14.0cm}{!} { \includegraphics{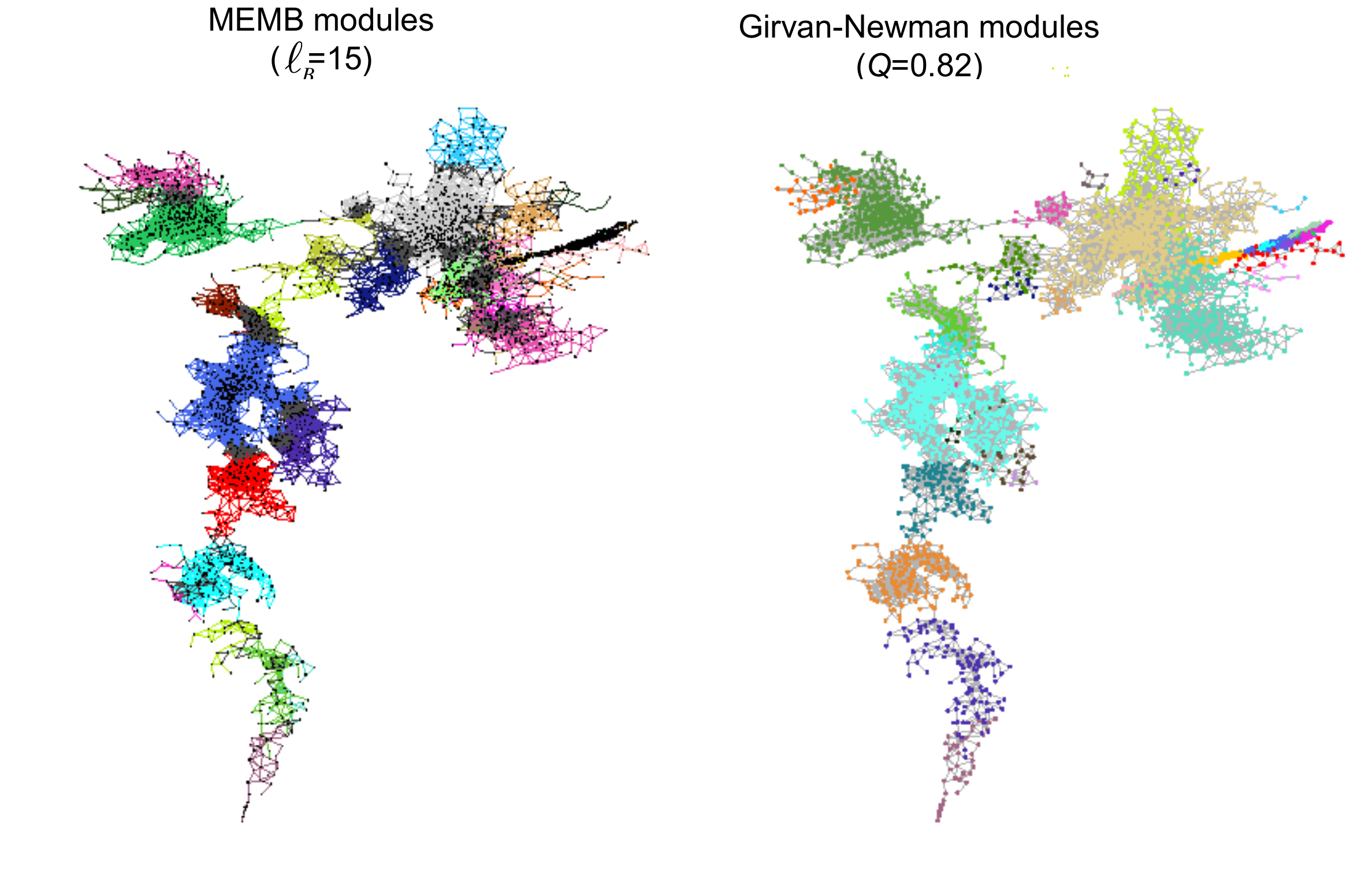}}}
  \caption{\label{si_newman}  {\bf Modular properties of the
      brain clusters}. Comparison between the partition provided by
    the MEMB method (at $\ell_B=15$) with the corresponding partition
    using the Garvin-Newman method [20]. The modularity
    index from the Newman definition $Q$ is around 0.82, as found by
    the latter method. Both methods yield similar sub-modules.}
\end{figure}

For the brain clusters we found an average modularity value of
$Q=0.82$. This is an indication of strong modularity within each
cluster.  A direct comparison between MEMB and the Girvan-Newman
method shows that they result in quite similar partitions.  We
calculated that 92\% of the total links belong within a given module
in both methods.  A visual comparison is shown in
Fig.~\ref{si_newman}. The maximization of modularity verifies the
modular character of the clusters.  The use of the MEMB, though,
provides us with the extra advantage of modifying the scale at which
we observe the modules to determine whether the modular structure is
scale-invariant, i.e., if it is composed of modules inside modules.

\begin{figure}
\centerline{ \resizebox{14.0cm}{!} { \includegraphics{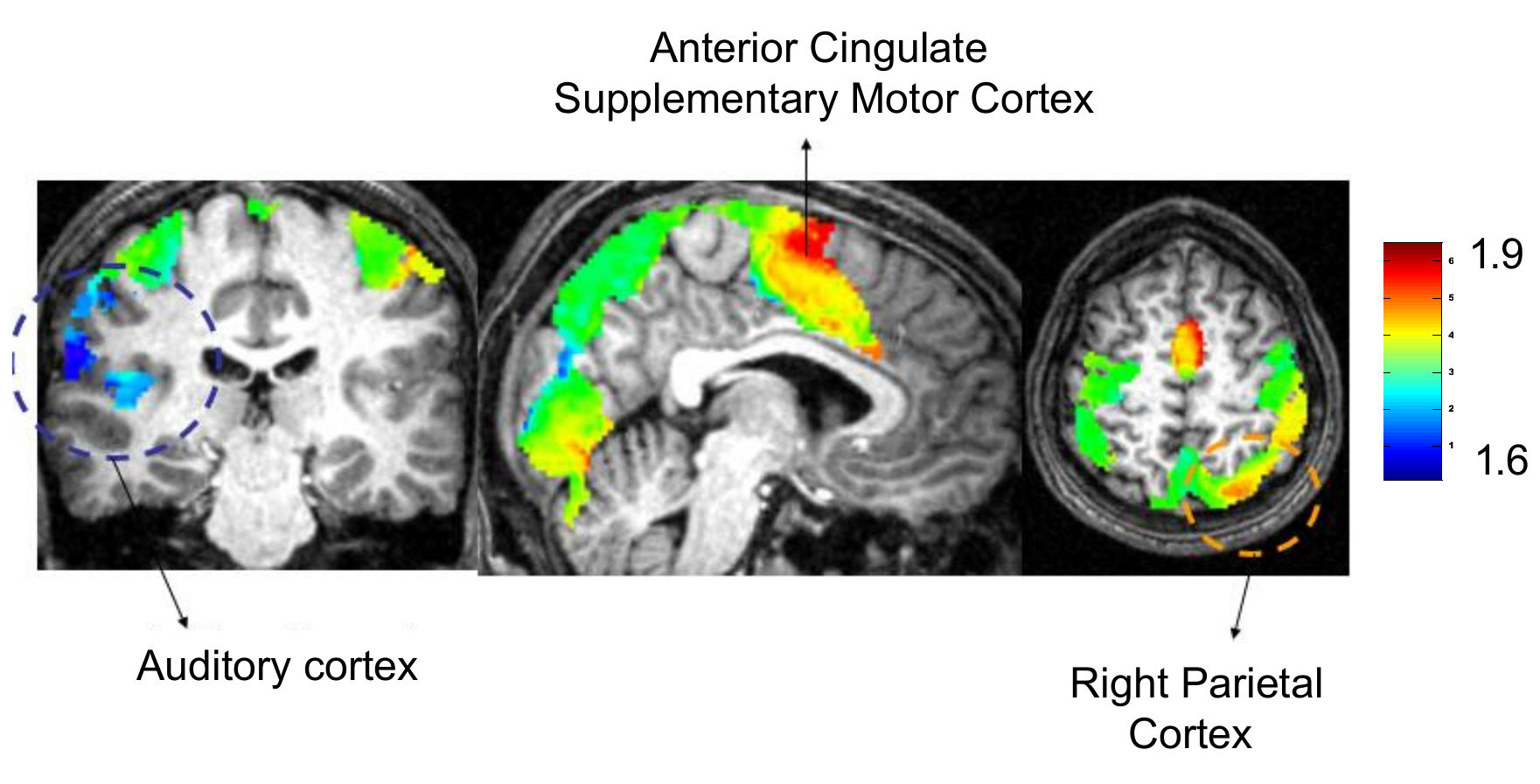}}}
  \caption{\label{fig5} {\bf Topographical map of module
      fractality}.  For each voxel, we calculate the average fractal
    dimension of the clusters to which it belongs, considering only
    voxels which form part of a cluster for at least eight subjects,
    to assure that mean values are not heavily determined by
    individual contributions. While the average over all clusters is
    $d_B=1.9 \pm 0.1$, the dimension of each cluster exhibits small
    variations around this value which allows us to identify
    consistent differences among them.  The clusters in the auditory
    cortex present the smaller fractal dimension $d_B$, while parietal
    and motor clusters show intermediate values of $d_B$. The right
    portion of the SMA and the right PPC were the clusters with the
    higher fractal dimension. }
  % MS Yes we need the colormap, I will send it soon.
\end{figure}

\end{document}